%% file: main.tex
\documentclass[10pt,a4paper]{article}
\input{preamble}
\input{definitions}
\input{metadata}
\begin{document}
\maketitle
\input{sections/abstract}
\input{sections/introduction}
\input{sections/data}

\input{sections/workflow}
\input{sections/results}

\input{sections/limitations}
\input{sections/conclusions}
\input{sections/acknowledgements}
\bibliographystyle{unsrt-doi}
\setlength{\multicolsep}{1pt}
\begin{multicols}{2}
\bibliography{references,references_heavyion}
\end{multicols}

\end{document}

%% file: preamble.tex
\usepackage[T1]{fontenc}
\usepackage{lmodern,microtype}
\usepackage[a4paper,left=12mm,right=12mm,top=12mm,bottom=12mm,headheight=12pt,headsep=3mm,footskip=6mm]{geometry}
\usepackage{amsmath,amssymb,booktabs,tabularx,array,graphicx,longtable}
\usepackage{subcaption,float,placeins,xcolor,fancyhdr,etoolbox,indentfirst}
\usepackage[colorlinks=true,linkcolor=blue!45!black,citecolor=blue!45!black,urlcolor=blue!45!black]{hyperref}
\hypersetup{pdftitle={AI-Assisted Analysis of CMS Open Data: Prompt J/psi Cross-Section Measurements in pp and Peripheral PbPb Collisions at sqrt(s)=7 TeV and sqrt(sNN)=2.76 TeV},pdfauthor={Zhipeng Cui, Ziyu Wang, Gerry Bauer, Kai Yi}}
\newcommand{\rev}[1]{{#1}}
\newcommand{\revb}[1]{{#1}}
\graphicspath{{figures/}}
\makeatletter
\patchcmd{\@maketitle}
  {\begin{tabular}[t]{c}\@author\end{tabular}}
  {\begin{minipage}{\textwidth}\centering\@author\end{minipage}}
  {}{\PackageError{inclusive-jpsi}{Author layout patch failed}{Check the document-class title layout.}}
\makeatother

\usepackage{titlesec,multicol}
\titleformat{\section}{\normalfont\large\bfseries}{\thesection}{.6em}{}
\titleformat{\subsection}{\normalfont\normalsize\bfseries}{\thesubsection}{.6em}{}
\titleformat{\subsubsection}[runin]{\normalfont\normalsize\bfseries}{\thesubsubsection}{.6em}{}[.]
\titlespacing{\section}{0pt}{6pt plus 1pt}{3pt}
\titlespacing{\subsection}{0pt}{5pt plus 1pt}{2pt}
\titlespacing{\subsubsection}{0pt}{3pt plus 1pt}{.65em}
\renewcommand{\arraystretch}{1.06}
\newsavebox{\CompactTableBox}
\newcommand{\FitTable}[1]{\sbox{\CompactTableBox}{#1}\ifdim\wd\CompactTableBox>\linewidth\resizebox{\linewidth}{!}{\usebox{\CompactTableBox}}\else\usebox{\CompactTableBox}\fi}
\makeatletter
\patchcmd{\@maketitle}{\vskip 2em}{\vskip -.7em}{}{}
\patchcmd{\@maketitle}{\vskip 1.5em}{\vskip .5em}{}{}
\patchcmd{\@maketitle}{\vskip 1em}{\vskip .3em}{}{}
\AtBeginEnvironment{thebibliography}{\fontsize{7.5}{8.5}\selectfont\setlength{\itemsep}{0pt}\setlength{\parskip}{0pt}}
\apptocmd{\thebibliography}{\setlength{\itemsep}{0pt}\setlength{\parsep}{0pt}}{}{}
\makeatother

%% file: definitions.tex
\newcommand{\jpsi}{J/\psi}
\newcommand{\pt}{p_{\mathrm T}}
\newcommand{\br}{\mathcal B(\jpsi\to\mu^+\mu^-)}
\newcommand{\xsec}{\br\,d^2\sigma_{\mathrm{prompt}}/(d\pt\,dy)}
\newcommand{\lxy}{\ell_{xy}}

%% file: metadata.tex
\title{AI-Assisted Analysis of CMS Open Data:\\
\revb{Prompt $J/\psi$ Cross-Section Measurements in $pp$ and Peripheral PbPb Collisions} at
$\sqrt{s}=7\,\mathrm{TeV}$ and
$\sqrt{s_{\mathrm{NN}}}=2.76\,\mathrm{TeV}$}

\author{%
  Zhipeng Cui\textsuperscript{1},
  Ziyu Wang\textsuperscript{1},
  Gerry Bauer\textsuperscript{1},
  Kai Yi\textsuperscript{1,*}\\[0.5em]
  {\small
    Institute of Physics Frontiers and Interdisciplinary Sciences,\\
    School of Physics and Technology, Ministry of Education Key Laboratory of\\
    Numerical Simulation for Large Scale Complex Systems (NSLSCS),\\
    Nanjing Normal University, Nanjing 210023, China
  }\\[0.5em]
  {\small
    *Corresponding author: \texttt{kai.yi@njnu.edu.cn}
  }%
}

\date{12 November, 2026}

%% file: sections/abstract.tex
\begin{abstract}

We investigate the use of an artificial-intelligence (AI) coding agent, supervised by human physicists, for collider physics analyses using CMS Open Data. OpenAI Codex performs dimuon event selection, two-dimensional invariant-mass and decay-length fits, Monte Carlo detector simulations, and systematic uncertainty evaluation after acquiring physics and experimental knowledge from published studies and external references, with human review at each stage. The workflow is tested on two CMS datasets, with custom full-detector simulations used to determine acceptance and reconstruction efficiencies. In proton--proton collisions at \(\sqrt{s}=7~\mathrm{TeV}\), prompt \(J/\psi\) differential cross sections are measured versus transverse momentum and rapidity. The ratios of remeasured to published CMS cross sections have a mean of 1.000 and an RMS spread of 0.215 across 39 bins. In peripheral (50--100\% centrality) PbPb collisions at \(\sqrt{s_{\mathrm{NN}}}=2.76~\mathrm{TeV}\), the corresponding prompt \(J/\psi\) fiducial cross-section ratios have a mean of 1.158 and an RMS spread of 0.225 across six intervals. We identify agent failure modes and demonstrate the importance of structured human review in ensuring reliable results.
\end{abstract}

%% file: sections/introduction.tex
\section{Introduction}

The public release of experimental data from high-energy, nuclear, and astroparticle physics experiments, including the Large Hadron Collider (LHC)~\cite{CERNOpenDataOverview2024}, RHIC~\cite{PHENIXOpenData2021}, Belle~\cite{Nishida2017BLab}, and IceCube~\cite{IceCubeOpenData2021}, enables independent analyses and reproduction of published measurements. Reliable reanalysis, however, requires not only event selection and background modeling, but also detector-efficiency corrections, systematic uncertainty evaluation, and validation. These challenges are particularly demanding across different collision systems, such as proton--proton (\(pp\)) and heavy-ion (PbPb) collisions, especially when dedicated signal Monte Carlo (MC) samples or precomputed trigger efficiencies are unavailable and legacy software and full-detector simulations must be configured.

Recent advances in large language models (LLMs) have opened new possibilities for scientific computing. Beyond code generation, LLM-based agents can execute programs, diagnose errors, and iteratively refine code. Emerging applications in high-energy physics (HEP) have demonstrated their potential in analyses of LEP and LHC Open Datasets, covering tasks from event selection and background estimation to uncertainty evaluation and statistical inference~\cite{GendreauDistler2025,Moreno2026}. Nevertheless, successful code execution alone does not ensure reliable physics results. Agents must correctly interpret analysis conventions, detector conditions, and statistical procedures, while their outputs require validation by physicists.

In this work, we investigate an AI-assisted workflow for CMS Open Data analysis using OpenAI's Codex coding agent~\cite{Chen2021Codex,OpenAICodexCLI}. Unlike conventional machine-learning methods, such as boosted decision trees~\cite{Roe2005BDT} and neural networks used for event classification or particle identification, our approach employs an LLM agent to implement and execute the analysis. Codex assists with data inspection, trigger and reprocessing checks, event selection, fitting, detector simulation, and systematic uncertainty evaluation. Human physicists provide domain knowledge and reference materials, define tasks, and validate intermediate and final results. The workflow incorporates published quarkonium studies, structured review checkpoints, and procedures for identifying and addressing agent failure modes.

Charmonium production provides a suitable benchmark for this approach, probing heavy-quark production and quarkonium formation in \(pp\) collisions~\cite{Bodwin1995,Brambilla2011,Cui2026}, as well as color screening and medium effects in heavy-ion collisions~\cite{CMSHI2011,CMSHI2011Jpsi}. As an initial validation, we analyze two CMS Open Data samples with corresponding published measurements: 2010 \(pp\) collisions at \(\sqrt{s}=7~\mathrm{TeV}\) and 2011 PbPb collisions at \(\sqrt{s_{\mathrm{NN}}}=2.76~\mathrm{TeV}\)~\cite{CERNOpenDataPortal,CMS2010Lumi,CMSHI2011OpenData,CMS2010,CMS2010Tables,CMSHI2011Jpsi}. These reference results allow us to assess the reliability of the workflow and establish a basis for future analyses without existing measurements for comparison. Prompt and nonprompt \(J/\psi\) mesons are separated using two-dimensional maximum-likelihood fits to the dimuon invariant mass and pseudo-proper decay length. The cross-section measurements require corrections for acceptance, reconstruction and trigger efficiencies, bin migration, and systematic uncertainties. Since dedicated reconstructed charmonium signal simulations are unavailable in the selected data releases, we generate them using Pythia6 interfaced with Geant4 for \(pp\) collisions, and HYDJET background generation followed by signal embedding for PbPb collisions. The \(pp\) analysis measures prompt \(J/\psi\) double-differential cross sections in 39 kinematic bins, while the peripheral PbPb analysis measures fiducial cross sections in six matched kinematic and centrality intervals.

The paper is organized as follows. Section~\ref{sec:data} describes the CMS detector, data samples, and simulation requirements. Section~\ref{sec:methods} presents the AI-assisted workflow architecture, the five analysis stages including uncertainty evaluation, and their implementation for the $pp$ and peripheral PbPb samples. Section~\ref{sec:results} presents the fit results, cross-section measurements, and comparisons with published CMS results. Section~\ref{sec:discussion} discusses agent failure modes, human oversight, and the remaining limitations and opportunities for improvement. Section~\ref{sec:conclusions} summarizes the findings and outlines future applications.

%% file: sections/data.tex
\section{Data samples}
\label{sec:data}

The collision data were recorded by the CMS detector~\cite{CMSDetector2008} at the CERN LHC. The central component of CMS is a superconducting solenoid with an internal diameter of $6\;\mathrm{m}$, providing an axial magnetic field of $3.8\;\mathrm{T}$. Inside the solenoid, a silicon pixel and strip tracker measures charged-particle trajectories within $|\eta|<2.5$ and provides the momentum and vertex information needed to reconstruct dimuon decays. Electromagnetic and hadronic calorimeters surround the tracker. The muon system, located outside the solenoid in the steel flux-return yoke, covers $|\eta|<2.4$ and consists of drift tubes in the barrel, cathode strip chambers in the endcaps, and resistive plate chambers in both regions. Muons are reconstructed by matching tracks in the silicon tracker to track segments in the muon chambers, and oppositely charged muon pairs are combined to form dimuon candidates.

Both collision datasets analyzed in this work were obtained from the CERN Open Data Portal~\cite{CERNOpenDataPortal,CMS2010Lumi,CMSHI2011OpenData}. They comprise two CMS dimuon samples from different collision systems: proton--proton ($pp$) collisions recorded in 2010 at $\sqrt{s}=7~\mathrm{TeV}$ and lead--lead (PbPb) collisions recorded in 2011 at a nucleon--nucleon centre-of-mass energy of $\sqrt{s_{\mathrm{NN}}}=2.76~\mathrm{TeV}$. The PbPb data were released in the CMS \texttt{RECO} format~\cite{CMSHI2011OpenData}. Table~\ref{tab:datasets} summarizes the collision energies, effective integrated luminosities $\mathcal{L}_{\mathrm{int}}$, open-data record identifiers, simulation configurations, and target observables and kinematic ranges.

The effective integrated luminosities listed in Table~\ref{tab:datasets} correspond to the certified luminosity accumulated while the selected unprescaled dimuon trigger was active. This distinction is important because high-rate dimuon triggers were often prescaled or active during only part of a data-taking period. For the 2010 $pp$ sample, selecting a single unprescaled dimuon trigger path gives an effective integrated luminosity of $3.04~\mathrm{pb}^{-1}$, compared with the total of $35.0~\mathrm{pb}^{-1}$ in the open-data release~\cite{CMS2010Lumi}. The published CMS measurement, by contrast, used multiple trigger periods and an integrated luminosity of $36.7~\mathrm{pb}^{-1}$~\cite{CMS2010}. For the 2011 PbPb sample, the official public reprocessing~\cite{CMSHI2011Reco} retained only peripheral collisions in the $50$--$100\%$ centrality interval. The corresponding effective integrated luminosity for the selected unprescaled dimuon trigger is $112~\mu\mathrm{b}^{-1}$. Effective integrated luminosities are quoted to three significant figures throughout.

Correcting the observed dimuon yields for geometric acceptance and detector reconstruction and trigger efficiencies requires simulated signal samples. However, the CERN Open Data Portal does not provide dedicated reconstructed charmonium simulations for either dataset. The AI agent therefore configured and executed custom full-detector simulations using containerized legacy CMS software releases and Geant4, as summarized in Table~\ref{tab:datasets}. From the perspective of evaluating the AI-assisted workflow, the two samples provide complementary tests:

\begin{itemize}
\item \textbf{Proton--proton cross-section extraction and custom detector simulation (2010 $pp$ at $\sqrt{s}=7~\mathrm{TeV}$):} Assesses the human--agent team's ability to perform two-dimensional invariant-mass and decay-length fits in the 44 $(p_{\mathrm{T}}, |y|)$ bins of the published measurement, and to generate a dedicated full-detector simulation of 238,180 unpolarized prompt $J/\psi\to\mu^+\mu^-$ decays for acceptance and efficiency corrections.

\item \textbf{Heavy-ion centrality auditing and signal-embedding simulation (2011 PbPb at $\sqrt{s_{\mathrm{NN}}}=2.76~\mathrm{TeV}$):} Assesses the workflow's ability to identify the $50$--$100\%$ peripheral centrality restriction in the public data release and reproduce the six corresponding prompt $J/\psi$ cross-section intervals in Ref.~\cite{CMSHI2011Jpsi}. The simulation involves embedding simulated $J/\psi\to\mu^+\mu^-$ decays into simulated minimum-bias PbPb background events.
\end{itemize}
\input{tables/datasets}

%% file: tables/datasets.tex
\begin{table}[htbp]
\centering\small
\caption{CMS Open Data samples, effective unprescaled trigger integrated luminosities $\mathcal{L}_{\mathrm{int}}$, dedicated signal Monte Carlo (MC) simulations, and analysis coverage~\cite{CERNOpenDataPortal,CMS2010Lumi,CMSHI2011OpenData,CMSHI2011Reco}. The AI workflow generated the signal MC using containerized legacy CMS simulation and reconstruction software. CERN record numbers identify the corresponding datasets in the CERN Open Data Portal.}
\label{tab:datasets}
\setlength{\tabcolsep}{6pt}
\renewcommand{\arraystretch}{1.18}
\begin{tabularx}{\linewidth}{@{}>{\raggedright\arraybackslash}p{0.21\linewidth}*{2}{>{\raggedright\arraybackslash}X}@{}}
\toprule
\revb{Sample information} & \revb{2010 $pp$} & \revb{2011 PbPb} \\
\midrule
Collision energy & $\sqrt{s}=7~\mathrm{TeV}$ & $\sqrt{s_{\mathrm{NN}}}=2.76~\mathrm{TeV}$ \\
Effective $\mathcal{L}_{\mathrm{int}}$ & $3.04~\mathrm{pb}^{-1}$ & $112~\mu\mathrm{b}^{-1}$ \\
Open Data sample & Dimuon data & PbPb dimuon \texttt{RECO} data \\
CERN record & 75, 76, 31109 & 14012 \\
Centrality & --- & 50--100\% \\
Software version & \texttt{CMSSW\_4\_2\_8} & \texttt{CMSSW\_4\_4\_7} \\
\addlinespace[3pt]
Custom signal MC & Pythia6 + Geant4 full detector simulation & Pythia6 $J/\psi$ embedded in quenched HYDJET PbPb events \\
\addlinespace[3pt]
Target observable & Prompt $J/\psi$,& Peripheral prompt $J/\psi$ \\
Kinematic coverage & $p_{\mathrm{T}}<30~\mathrm{GeV}$, $|y|<2.4$ & $|y|<2.4$ \\
\bottomrule
\end{tabularx}
\end{table}

%% file: sections/workflow.tex
\section{AI-assisted analysis workflow}\label{sec:methods}
The physics analysis follows standard experimental methods: kinematic and vertex selections, unbinned one- and two-dimensional maximum-likelihood fits, and Monte Carlo (MC) efficiency corrections for geometric acceptance, reconstruction and trigger efficiencies, and bin-to-bin migration. The AI agent operates at the software and workflow level, writing, compiling, executing, and debugging the analysis and simulation code under the supervision of human physicists. This section describes the agent and its operating structure (Section~\ref{sec:aimethod}), the five-stage workflow (Section~\ref{sec:ailifecycle}), and how the workflow was applied to the two analyses (Section~\ref{sec:twoanalyses}).

\subsection{AI architecture}\label{sec:aimethod}

\paragraph{Three-layer architecture.}
A broad instruction such as ``measure the CMS $J/\psi$ cross section'' leaves many essential analysis choices unspecified. Without further guidance, an agent might select an incorrect trigger, use inappropriate kinematic boundaries, or overlook restrictions imposed by the public data reprocessing. To address these challenges, we organize the workflow into three interacting layers (Fig.~\ref{fig:aiworkflow}):

\begin{enumerate}
\setlength{\itemsep}{0.5pt}\setlength{\parskip}{0pt}
\item \textbf{Knowledge learning layer:} This layer provides the physics and experimental knowledge needed to guide the analysis. The agent starts with a local collection of published CMS quarkonium papers~\cite{CMS2010,CMSHI2011,CMSHI2011Jpsi,CMSHI2011Calibration}, machine-readable numerical tables~\cite{CMS2010Tables}, and records of open-data luminosity and reprocessing~\cite{CMS2010Lumi,CMSHI2011OpenData,CMSHI2011Reco}. It also actively searches external references, including the Particle Data Group (PDG) and related publications, to acquire additional information and cross-check the available material. From these sources, it establishes the $p_{\mathrm{T}}$, rapidity ($y$), and centrality bin boundaries; identifies the appropriate trigger paths and vertex-quality requirements; and determines the signal and background models. When a publication describes a fit function only qualitatively, the agent implements a standard RooFit parameterization and tests alternative forms to assess model dependence.

\item \textbf{Agentic coding and orchestration layer:} This layer develops and manages the software needed for the analysis and the production of simulated samples. The agent writes C++/ROOT and Python programs for event processing, dimuon candidate reconstruction, likelihood fitting, and efficiency calculations. It also configures the Monte Carlo generation and detector-simulation workflows using Pythia6, HYDJET, and Geant4. A structured project directory keeps the data, simulation configurations, analysis code, diagnostic plots, and validation logs organized. The agent compiles and tests the software, runs small-scale jobs interactively, and submits large-scale simulation and analysis productions to the computing farm.

\item \textbf{Agentic analysis layer:} This layer applies the developed software to the data and simulated samples to obtain the physics results. Guided by the acquired domain knowledge, the agent selects the appropriate trigger paths and applies event- and muon-selection criteria to construct dimuon candidates. It then performs invariant-mass and decay-length fits to separate prompt and nonprompt contributions, determines acceptance and reconstruction and trigger efficiencies, and extracts the cross sections. Systematic uncertainties are evaluated by varying the analysis selections, fit models, and other relevant assumptions. Fit diagnostics, numerical results, and comparison plots are documented at each stage for review and validation by human physicists.
\end{enumerate}

\input{sections/compact_figures/aiworkflow}

\subsection{Workflow}\label{sec:ailifecycle}

Rather than assigning the entire measurement through a single open-ended prompt, we divide the analysis into five sequential stages: learning, event selection, information extraction, MC correction, and systematic uncertainty evaluation. At each stage, the physicist provides a focused instruction, and the agent develops and executes the required software. Importantly, the agent is not restricted to the information supplied with the instruction. When essential details are missing, it can actively search the public web for additional information, consulting authoritative sources such as the Particle Data Group (PDG), relevant research publications, and technical documentation. It uses these sources to fill knowledge gaps and guide the implementation, while recording the references for human verification. Each stage ends at a mandatory review checkpoint, where the physicist examines the numerical results and diagnostic plots before the workflow proceeds. This combination of active information retrieval, iterative implementation, and human review helps prevent errors from propagating through the analysis.

\paragraph{Stage 1: Learning.}
Before processing events, the agent studies the reference publication, inspects the open-data files, and examines the corresponding luminosity records. It also searches for additional information whenever the available references do not fully specify the analysis procedure. A representative instruction is:

\emph{``Read the 2010 CMS $J/\psi$ paper and the public tables in our reference collection. Inspect the 2010 dimuon open-data files and luminosity record. Identify the unprescaled dimuon trigger path, calculate its certified integrated luminosity, and summarize the $p_{\mathrm{T}}$ and $|y|$ bin boundaries and vertex requirements. Search for additional documentation if any of these details are unclear.''}

The purpose of this stage is to establish the analysis conventions and verify that they are compatible with the actual data before writing the analysis code. The agent summarizes the key definitions for human review, including the rapidity and centrality intervals, the reference point used to calculate the decay length, and whether the reported cross section includes the dimuon branching fraction.

\paragraph{Stage 2: Event selection.}
The agent converts the released event files into flat ntuples of dimuon candidates and applies the selections specified in the reference publication. It also audits the data release for restrictions that could affect comparisons with published results. The following general instruction is adapted to each dataset:

\emph{``Loop over the released event files and create one ntuple entry for each opposite-sign dimuon candidate. Store the muon kinematics and quality variables, vertex-fit probability, trigger decisions, and event luminosity-section information. Apply the muon, vertex, and kinematic selections from the reference paper. Identify the available trigger paths and their prescale status, then use the certified luminosity table to calculate the effective integrated luminosity for the selected path. For heavy-ion data, inspect the reprocessing configuration and identify the retained centrality range. Report the number of events and candidates after each selection.''}

The resulting cut-flow table is compared with the published analysis at the control gate. This audit is essential because limitations of the public data can affect the validity of the measurement. For example, failure to account for trigger prescales can lead to an incorrect effective luminosity, while reprocessing filters can restrict the available phase space. A peripheral PbPb sample covering $50$--$100\%$ centrality, for instance, cannot be directly compared with a published result for $0$--$20\%$ central collisions. The dataset-specific selections are described in Section~\ref{sec:twoanalyses}.

\paragraph{Stage 3: Information extraction.}\label{sec:observables}
Prompt and nonprompt $J/\psi$ mesons are separated using a simultaneous analysis of the dimuon invariant mass and decay-length distributions. The agent first determines the observable definition and fitting procedure from the reference publication. A representative instruction is:

\emph{``Determine how the reference paper separates prompt and nonprompt $J/\psi$ mesons. Identify the exact definition of the decay-length variable, including its reference point (beam spot or primary vertex) and the mass used in its calculation. Calculate the variable and its per-candidate uncertainty from the vertex fit. Plot its distribution in one kinematic bin and identify the contributions from prompt candidates, nonprompt candidates, and detector resolution.''}

The two muons in each candidate are fitted to a common vertex. The prompt and nonprompt contributions are then separated statistically using a joint unbinned maximum-likelihood fit to the dimuon invariant mass $m_{\mu^+\mu^-}$ and the signed pseudo-proper decay length $\lxy$, defined as
\begin{equation}
 \lxy=\frac{M_{\mathrm{def}}}{\pt^2}
 (\boldsymbol v_{xy}-\boldsymbol v_{\mathrm{ref},xy})\cdot\boldsymbol p_{\mathrm T}.
 \label{eq:lifetime}
\end{equation}
Here, $\boldsymbol v_{xy}$ is the reconstructed transverse position vector of the dimuon decay vertex, and $\boldsymbol v_{\mathrm{ref},xy}$ is the transverse position vector of the collision reference point. The latter differs between the two collision systems and must follow the corresponding published definition. The 2010 $pp$ analysis~\cite{CMS2010} uses the transverse beam-spot position, representing the center of the luminous region, whereas the 2011 PbPb analysis~\cite{CMSHI2011Jpsi} uses the event-by-event primary vertex reconstructed from charged-particle tracks. These definitions are not interchangeable. The vector $\boldsymbol p_{\mathrm T}$ is the dimuon transverse momentum, with magnitude $\pt=|\boldsymbol p_{\mathrm T}|$, and $M_{\mathrm{def}}$ is the reconstructed dimuon invariant mass used in the corresponding publication to convert the transverse flight distance into a proper decay length.

Negative values of $\lxy$ arise from finite detector resolution and help constrain the prompt resolution model. The extended maximum-likelihood fit includes prompt signal, nonprompt signal, and continuum background components. The signal mass distribution accounts for the radiative low-mass tail using a Crystal Ball plus Gaussian function, while the $\lxy$ resolution is parameterized using the per-candidate decay-length uncertainty.

A representative implementation instruction is:

\emph{``Implement the unbinned two-dimensional maximum-likelihood fit to dimuon mass and pseudo-proper decay length $\ell_{J/\psi}$ in each 2010 kinematic bin, following the paper's Crystal Ball plus Gaussian signal and resolution models and scaling $\ell_{J/\psi}$ by the reconstructed dimuon mass. Check that the covariance matrix is positive-definite. Compare the fitted mass mean, resolution width, and nonprompt decay-length slope with the published values. Save the mass and decay-length projections, including pull distributions, for review.''}

At the control gate, every nominal and alternative-model fit must converge with a positive-definite covariance matrix and an estimated distance to the minimum satisfying $\mathrm{EDM}<10^{-2}$. The mass and decay-length projections, together with their pull distributions, are also inspected visually.

\paragraph{Stage 4: MC correction.}\label{sec:2012case}
This stage converts the fitted signal yields into cross sections by correcting for geometric acceptance, reconstruction, vertexing and trigger efficiencies, and bin-to-bin migration, as described by Eq.~\eqref{eq:response}. The agent first studies the correction procedure in the reference publication, including the fiducial muon acceptance, efficiency factorization, assumed $J/\psi$ polarization, and data-driven efficiency calibrations. It then checks the open-data portal for a matching signal simulation. A representative instruction is:

\emph{``Summarize the acceptance definition, efficiency components, and calibrations used in the reference paper. Check whether a signal simulation matching our data-taking period is available on the portal, and compare its software release, detector conditions, and generator filters with those of our data. If no suitable sample exists, configure and run a full detector simulation using the appropriate legacy software release and generator settings. Emulate the selected trigger path and calculate acceptance and efficiency in the analysis bins using the same selections as in data.''}

\begin{figure}[htbp]
\centering
\includegraphics[width=0.96\linewidth]{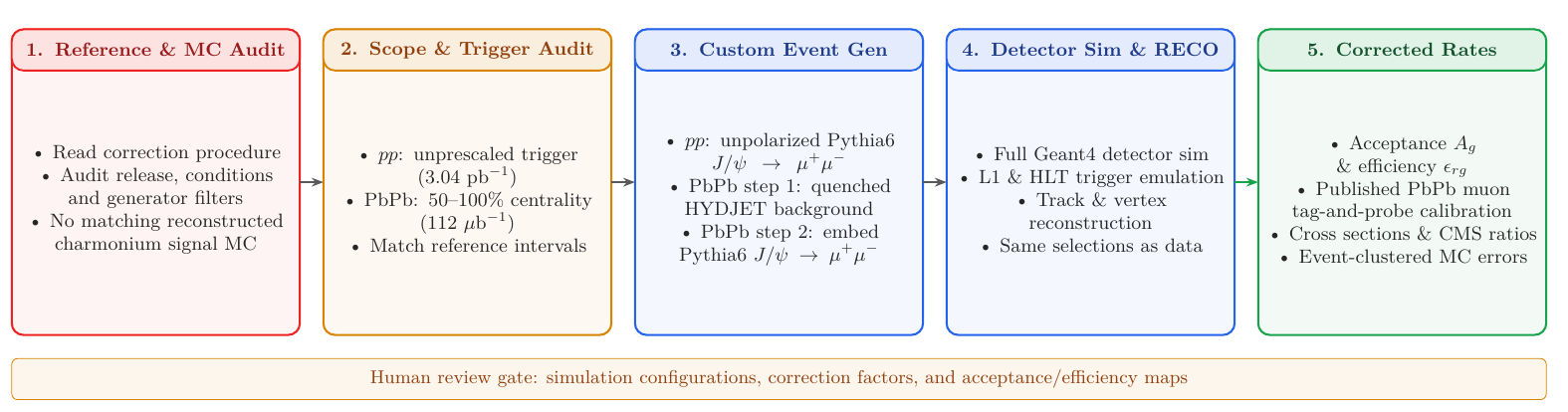}
\caption{Reference-guided archive audit and custom full-simulation pipeline for the 2010 7~TeV $pp$ and 2011 2.76~TeV peripheral PbPb analyses. The PbPb response includes the published single-muon efficiency calibration; simulation configurations, correction factors, and acceptance and efficiency maps undergo human review.}
\label{fig:cms2012recovery}
\end{figure}

We deliberately select two CMS Open Data samples without publicly available reconstructed charmonium signal MC samples to test the agent's ability to perform full-detector simulations from scratch. The agent must therefore configure and execute dedicated full-detector simulations, as illustrated in Fig.~\ref{fig:cms2012recovery}. The procedures differ between the two collision systems. For $pp$, the agent audits the run-by-run trigger records to identify a single unprescaled trigger path whose Level-1 and high-level trigger filter logic can be reproduced with the legacy software. It then generates Pythia6 events and processes them through the Geant4 detector simulation. For PbPb, the agent first examines the official reprocessing configuration to establish the retained centrality range. It then builds a two-step simulation campaign: minimum-bias PbPb background events are generated with HYDJET, and Pythia6 quarkonium decays are embedded into these events before Geant4 simulation, trigger emulation, and track reconstruction.

When the reference paper's efficiency calibration (such as single-muon tag-and-probe results) is published as a parameterization, the agent applies it to the simulated response; the simulation, the correction factors, and the resulting acceptance and efficiency maps are reviewed at the control gate.

The primary observable is the double-differential prompt cross section multiplied by the dimuon branching fraction, $\xsec$. For peripheral PbPb collisions, we also consider the fiducial cross section $\sigma_{\mathrm{fid}}$ and nuclear-overlap-scaled yield density $\rho$. These observables are evaluated in bins of transverse momentum $\pt$ and rapidity $y$.

Let $g$ and $r$ denote the generated and reconstructed kinematic bins, respectively. The fitted prompt signal yield $Y_r$ in reconstructed bin $r$ is related to the bin-averaged cross section times branching fraction $x_g=(\xsec)_g$ by
\begin{equation}
 Y_r=\mathcal L\sum_g\mathcal R_{rg}\,x_g,
 \qquad
 \mathcal R_{rg}=\Delta p_{\mathrm T,g}\,\Delta y_g\,A_g\,\epsilon_{rg}.
 \label{eq:response}
\end{equation}
Here, $\mathcal L$ is the effective integrated luminosity, and $\Delta p_{\mathrm T,g}$ and $\Delta y_g$ are the generated-bin widths, with $\Delta y=2\Delta|y|$ for symmetric rapidity intervals. The quantity $A_g$ denotes the geometric and kinematic acceptance for unpolarized $J/\psi\to\mu^+\mu^-$ decays, including final-state photon radiation. The factor $\epsilon_{rg}$ is the combined reconstruction, vertexing, and trigger efficiency for an accepted event generated in bin $g$ to be selected and reconstructed in bin $r$.

In both the 2010 $pp$ and 2011 PbPb analyses, the custom full-simulation samples are used to determine $A_g$ and the bin-by-bin efficiency correction. The latter relates selected reconstructed prompt candidates in bin $i$ to truth-fiducial prompt parents under the generator spectrum, thereby accounting for weighted bin-to-bin migration and feed-in from outside the fiducial boundaries. For the 2011 PbPb analysis, the simulated response is additionally corrected using the published single-muon tag-and-probe data-to-simulation efficiency factors~\cite{CMSHI2011Calibration}.

\paragraph{Stage 5: Systematic uncertainties.}\label{sec:systematics}
A key test of an AI coding agent is whether it can perform a physically sound uncertainty evaluation rather than simply report the statistical errors returned by Minuit (fit package). The agent identifies potential systematic sources using three complementary inputs: the uncertainty evaluation in the reference publication, the individual factors entering Eq.~\eqref{eq:response}, and additional public documentation or analysis notes when the published procedure lacks sufficient detail.

A representative instruction is:

\emph{``List all systematic uncertainty sources discussed in the reference paper. For each source, explain how it was evaluated and whether the procedure can be reproduced with our open data and simulation. If the paper does not provide sufficient detail, search public documentation for the standard procedure and cite the relevant source. Implement each evaluable source as a separate variation, tabulate its effect on the signal yield and cross section, and explicitly identify sources that cannot be evaluated.''}

The agent's proposed sources and supporting references are reviewed by the physicist, who determines which procedures are appropriate. The evaluation considers the statistical uncertainty on the fitted signal yield $Y_r$, the finite size of the simulated samples entering $A_g$ and $\epsilon_{rg}$, the sensitivity of the yields to signal, background, and resolution models, external efficiency calibrations, and the integrated luminosity $\mathcal L$.

An unguided agent can make several important mistakes, including treating multiple reconstructed candidates from the same simulated collision as statistically independent, adding correlated model variations in quadrature, and silently assigning zero uncertainty to sources for which no calibration is available. We address these problems through three rules enforced at the control gate:

\begin{itemize}
\setlength{\itemsep}{0.5pt}\setlength{\parskip}{0pt}
\item \textbf{Model sensitivity evaluated by refitting:} Each kinematic interval is refitted using alternative signal-shape, mass-background, lifetime-background, and resolution parameterizations. Only fits that converge with positive-definite covariance matrices are retained. For each model component, the largest absolute yield shift is taken as a conservative uncertainty envelope rather than adding variations of the same component in quadrature.

\item \textbf{Simulation statistics evaluated by event clustering:} Reconstructed signal candidates are grouped according to their parent simulated event, or the shared background event in embedded samples, when calculating the MC statistical uncertainty. This preserves correlations among multiple candidates originating from the same event.

\item \textbf{Unevaluated sources explicitly documented:} Systematic sources that cannot be validated, such as data-driven trigger scale factors, pileup reweighting, feed-down corrections, and detector-occupancy closure, are identified as unevaluated in table notes and figure captions. They are never silently assigned zero uncertainty.
\end{itemize}

The dataset-specific implementation of these procedures is described in Sections~\ref{sec:cms2010results} and~\ref{sec:cmsHI2011}.

\subsection{\texorpdfstring{Validation with CMS $pp$ and PbPb Open Data}{Validation with CMS pp and PbPb Open Data}}
\label{sec:twoanalyses}

We applied the workflow to a proton--proton ($pp$) analysis and a peripheral lead--lead (PbPb) analysis. This subsection describes their dataset-specific implementation; the measurements and comparisons with published CMS results are presented in Section~\ref{sec:results}.

\subsubsection{\texorpdfstring{Prompt $J/\psi$ production in $pp$ collisions at $\sqrt{s}=7~\mathrm{TeV}$ (2010)}{Prompt J/psi production in pp collisions at sqrt(s)=7 TeV (2010)}}
\label{sec:cms2010results}

We use the 2010 $pp$ dataset listed in Table~\ref{tab:datasets}~\cite{CERNOpenDataPortal}. The public archive provides neither dedicated signal simulations nor precomputed trigger efficiencies.

\paragraph{Event selection and trigger-luminosity audit.}
Following the published CMS analysis~\cite{CMS2010}, the agent selected opposite-sign dimuon pairs from 52,261,231 certified collision events, containing 499,058 raw pairs. Requirements included a common-vertex fit probability above $1\%$, retention of the candidate with the highest vertex-fit probability in each event, and a requirement that the two muons bend away from each other in the transverse magnetic field. The latter is expressed as $\phi(\mu^+)-\phi(\mu^-)<0$, wrapped to $(-\pi,\pi]$, and suppresses closely spaced pairs that share trigger hits in the barrel--endcap overlap region. The selection yields 23,338 candidates.

The per-event candidate arbitration follows Section~3 of Ref.~\cite{CMS2010}. After the vertex-probability and kinematic requirements, 43,294 pairs remain in 43,248 events; only 44 events ($0.102\%$) contain multiple pairs, so arbitration removes 46 pairs before the azimuthal bending requirement. The same procedure is applied to simulation, incorporating its selection loss into the detector response. Residual data-to-simulation differences in candidate competition and pileup are not included in the evaluated systematic uncertainties.

The published analysis used $36.7~\mathrm{pb}^{-1}$ across several dimuon trigger periods~\cite{CMS2010}. The corresponding data-driven tag-and-probe trigger calibrations are not available in machine-readable form. We therefore restrict the reanalysis to one unprescaled ``Tight'' double-muon trigger, retaining 21,804 unique events. Its Level-1 and high-level trigger logic can be emulated directly in simulation. Matching the active runs and luminosity sections to the certified luminosity table gives an effective integrated luminosity of $3.04~\mathrm{pb}^{-1}$.

This restriction avoids potential biases from an uncalibrated trigger mixture and enables a consistent simulation-based efficiency correction, but reduces the integrated luminosity by approximately $91\%$ relative to the published sample. The statistical uncertainties consequently increase by about $\sqrt{36.7/3.04}\approx3.5$, and five sparsely populated phase-space bins remain empty. Nevertheless, luminosity normalization and detector corrections permit an absolute cross-section comparison with the published CMS spectrum~\cite{CMS2010Tables}.

\paragraph{Custom simulation and bin-by-bin efficiency correction.}
The agent configured and ran a full-detector simulation of 238,180 unpolarized prompt $J/\psi\to\mu^+\mu^-$ decays using the legacy 2010 CMS software and Geant4 (Table~\ref{tab:datasets}). At generator level, it evaluated the acceptance $A_g$ in each $(p_{\mathrm{T}},|y|)$ bin for prompt $J/\psi$ mesons with $|y^{J/\psi}|<2.4$. Following Eq.~(1) of Ref.~\cite{CMS2010}, both decay muons must satisfy
\begin{equation}
|\eta^\mu| < 2.4, \qquad
p_{\mathrm{T}}^\mu >
\begin{cases}
4.0~\mathrm{GeV}, & |\eta^\mu| < 1.2, \\
3.3~\mathrm{GeV}, & 1.2 \le |\eta^\mu| < 2.4.
\end{cases}
\label{eq:cms2010_fiducial}
\end{equation}
The fully reconstructed simulation provides the bin-by-bin reconstruction, vertexing, and Tight-trigger response, relating selected reconstructed candidates in bin $i$ to truth-level fiducial prompt parents according to Eq.~\eqref{eq:response}. This accounts for weighted bin migration and feed-in from events generated outside the single-muon fiducial boundaries.

\paragraph{Systematic uncertainties.}
The integrated-luminosity uncertainty is $11\%$~\cite{CMS2010Lumi}. The fit-model uncertainty is evaluated using four alternatives: a pure Gaussian signal instead of the nominal Crystal Ball plus Gaussian, a linear mass background, a single-Gaussian prompt lifetime resolution, and a floating resolution scale factor. Monte Carlo statistical uncertainties are evaluated by clustering candidates according to their parent simulated collision event. Data-driven trigger scale factors, pileup reweighting, and $\chi_c/\psi(2S)$ feed-down corrections are unavailable and remain unevaluated.

\subsubsection{\texorpdfstring{Prompt $J/\psi$ production in peripheral PbPb collisions at $\sqrt{s_{\mathrm{NN}}}=2.76~\mathrm{TeV}$ (2011)}{Prompt J/psi production in peripheral PbPb collisions at sqrt(sNN)=2.76 TeV (2011)}}
\label{sec:cmsHI2011}

The PbPb analysis addresses two additional challenges: the restricted centrality coverage of the public data and the need to embed simulated quarkonium decays into heavy-ion background events for efficiency corrections.

\paragraph{Centrality audit and event selection.}
The public 2011 PbPb dimuon \texttt{RECO} dataset (Table~\ref{tab:datasets})~\cite{CMSHI2011OpenData} contains 7,000,799 certified collision events. By inspecting the official CMS reprocessing configuration~\cite{CMSHI2011Reco}, the agent established that the release had been filtered during reprocessing to retain only peripheral collisions in the $50$--$100\%$ centrality range.

The primary CMS charmonium-suppression analysis~\cite{CMSHI2011} reports prompt $\psi(2S)$-to-$J/\psi$ double ratios between PbPb and $pp$ collisions in the $0$--$20\%$, $20$--$40\%$, and $40$--$100\%$ centrality intervals. None is fully covered by the available data. Although the agent also performed inclusive two-peak mass fits in the $50$--$100\%$ sample, these uncorrected single ratios are not directly comparable to the published prompt double ratios. We therefore chose a directly comparable observable: efficiency-corrected peripheral prompt $J/\psi$ cross sections in the six $(p_{\mathrm{T}},|y|,\text{centrality})$ intervals of Ref.~\cite{CMSHI2011Jpsi}, which use the same $50$--$100\%$, $50$--$60\%$, and $60$--$100\%$ centrality ranges.

The selection applies hadronic collision, opposite-sign muon quality and kinematic acceptance, unprescaled dimuon trigger, and common-vertex fit probability requirements. After excluding invalid or conflicting luminosity sections and requiring an unprescaled selected high-level trigger (HLT) path, the effective integrated luminosity is $112~\mu\mathrm{b}^{-1}$. The initial selection yields 3,638 candidates across the midrapidity ($|y|<1.6$, $6.5<\pt<30~\mathrm{GeV}$) and forward ($1.6<|y|<2.4$, $3<\pt<30~\mathrm{GeV}$) regions. Applying the final mass--lifetime fit-window and per-event decay-length uncertainty requirements of Ref.~\cite{CMSHI2011Jpsi} leaves 1,625 unique pairs for the fits in Table~\ref{tab:cmsHI2011prompt}: 988 in $6.5<\pt<30~\mathrm{GeV}$ (469, 226, and 293 across the three rapidity intervals, or 556 and 432 across the two centrality intervals) and 637 in the disjoint low-$\pt$ interval $3<\pt<6.5~\mathrm{GeV}$.

\paragraph{Two-step HYDJET background simulation and $J/\psi$ signal embedding.}
Since no reconstructed PbPb charmonium simulation is available in the CERN Open Data Portal, the agent developed a two-step embedding workflow using the legacy 2011 CMS heavy-ion software and Geant4 (Table~\ref{tab:datasets}). First, it generated quenched minimum-bias PbPb background events at $\sqrt{s_{\mathrm{NN}}}=2.76~\mathrm{TeV}$ with HYDJET~\cite{Lokhtin2006HYDJET}, which combines hydrodynamic soft-particle production with hard scatterings and jet quenching. It then embedded Pythia6 prompt $J/\psi\to\mu^+\mu^-$ decays into these events and performed Geant4 detector simulation, Level-1 and high-level trigger emulation, and event reconstruction.

The CMS reconstruction ``scenario'' flag selects the charged-particle track-reconstruction sequence without modifying the simulated collision event. We used the standard proton--proton track-reconstruction sequence, matching the official reprocessing configuration of the released $50$--$100\%$ PbPb data (Record 3420~\cite{CMSHI2011Reco}). However, the open data were reprocessed from existing \texttt{RECO} products, whereas our simulation runs reconstruction from simulated \texttt{RAW} data. The transfer of external detector-efficiency calibrations to this embedded sample therefore remains unverified.

Multiple embedded candidates can share a HYDJET background event and are correlated through its track multiplicity. The agent accounted for this correlation by clustering candidates by their parent HYDJET event and using a cluster bootstrap to propagate Monte Carlo statistical uncertainties across the six peripheral intervals.

\paragraph{Joint two-dimensional prompt $J/\psi$ fits and automated handling of fit degeneracies.}
The agent performed joint two-dimensional unbinned maximum-likelihood fits to dimuon invariant mass and pseudo-proper decay length $\ell_{J/\psi}$ in the six peripheral intervals of Ref.~\cite{CMSHI2011Jpsi}, separating prompt $J/\psi$ mesons from nonprompt contributions from $b$-hadron decays. The nominal model uses a Crystal Ball plus Gaussian signal shape and an exponential decay-length parameterization. Alternative signal shapes, background polynomial orders, and lifetime-resolution models were tested; the fit-model uncertainty is the largest absolute yield shift among converged fits with positive-definite covariance matrices (Table~\ref{tab:cmsHI2011prompt}).

An automated covariance check identified a degeneracy in the $60$--$100\%$ centrality interval (432 pairs, marked by $\dagger$ in Table~\ref{tab:cmsHI2011prompt}). In the nominal fit, the negative-decay-length background amplitude converged to zero, leaving its decay constant unconstrained and producing a singular covariance matrix with an artificially small symmetric uncertainty. The agent fixed the vanishing amplitude to zero and evaluated asymmetric profile-likelihood uncertainties of $256.0^{+18.2}_{-17.6}$ events. This restored a well-defined covariance matrix while changing the central prompt yield by less than $0.01$ events.

\paragraph{Calibration transfer and cross-section observables.}
The fitted yields are used to determine the fiducial differential cross section
\[
\sigma_{\mathrm{fid}} \equiv
\mathcal{B}(J/\psi\to\mu^+\mu^-)
\frac{d\sigma_{\mathrm{fid}}}{dp_{\mathrm{T}}},
\]
in nb/GeV, integrated over the specified rapidity interval, and the nuclear-overlap-scaled differential yield
\[
\rho \equiv
\mathcal{B}(J/\psi\to\mu^+\mu^-)
\frac{dN_{\mathrm{PbPb}}/dp_{\mathrm{T}}}
{\langle T_{\mathrm{AA}}\rangle},
\]
in pb/GeV. The agent combined the HYDJET-embedded detector response with the published single-muon tag-and-probe efficiency parameterization~\cite{CMSHI2011Calibration}, propagating correlated two-muon efficiency variations across all six intervals.

\paragraph{Systematic uncertainties.}
The agent evaluated variations of the background polynomial, signal shape, and radiative-tail parameters in both the inclusive $\psi(2S)/J/\psi$ mass fits and the two-dimensional prompt $J/\psi$ fits. The uncertainty envelope is the largest yield shift among converged refits. In the two inclusive PbPb domains, the maximum shifts are $\pm14.1$ and $\pm10.6$ $J/\psi$ candidates, and $\pm1.5$ and $\pm2.2$ $\psi(2S)$ candidates.

Monte Carlo statistical uncertainties are evaluated by clustering candidates by their shared HYDJET background event. Correlated two-muon variations of the published single-muon tag-and-probe efficiencies~\cite{CMSHI2011Calibration} are propagated across the six intervals. Absolute heavy-ion tracking-efficiency and occupancy-closure corrections have not been established for the embedded simulation and remain unevaluated.

%% file: sections/compact_figures/aiworkflow.tex
\begin{figure}[htbp]
\centering
\includegraphics[width=0.96\linewidth]{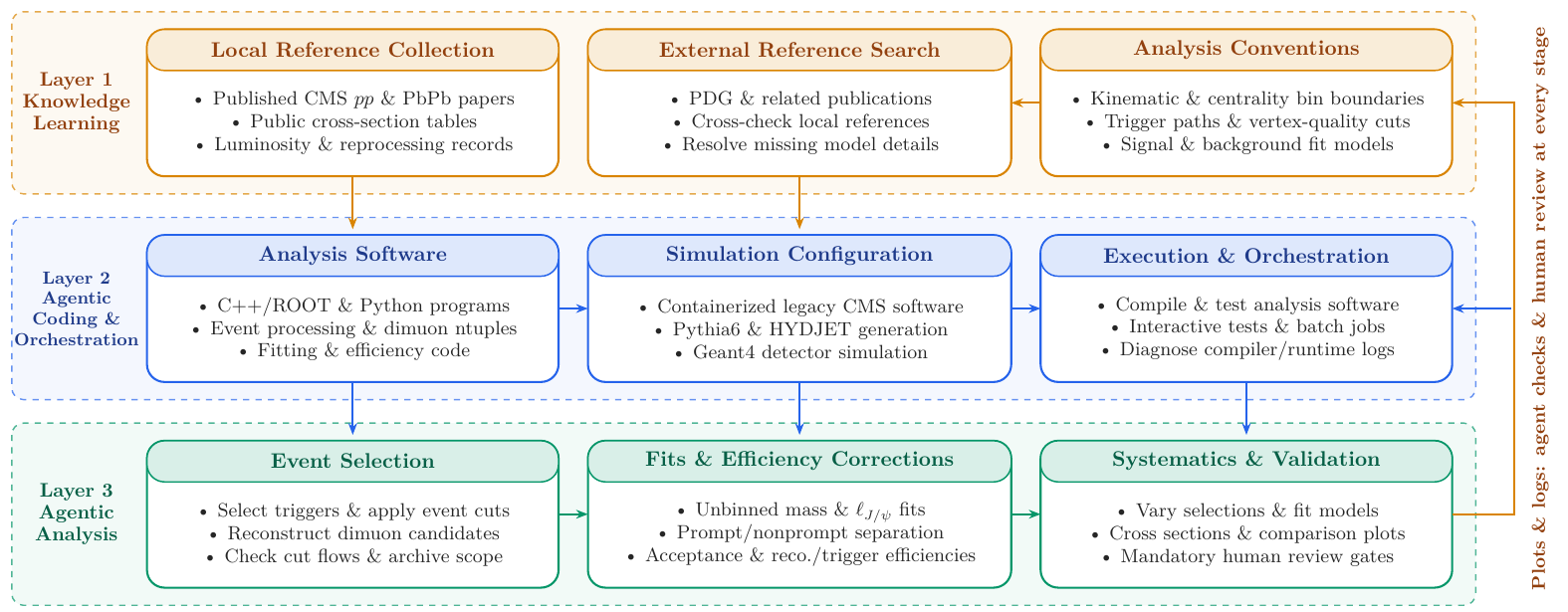}
\caption{Three-layer architecture of the AI-assisted analysis workflow: knowledge learning from local and external references at the top (Layer~1, amber); agentic coding and orchestration in the middle (Layer~2, blue); and agentic analysis at the bottom (Layer~3, green). A closed feedback loop returns plots and logs for agent self-checks and mandatory human review at every stage.}
\label{fig:aiworkflow}
\end{figure}

%% file: sections/results.tex
\section{Results and comparison with published measurements}
\label{sec:results}

\paragraph{Comparison procedure.}
\label{sec:validation}
\label{sec:validation_metrics}
To assess how well the AI-assisted reanalysis reproduces the published CMS measurements, we compare the prompt cross section times branching fraction in each kinematic bin $i$, denoted by $x_i$, with the corresponding published value, $x_i^{\mathrm{pub}}$.

\rev{The uncertainties in the two measurements are correlated because the reanalysis uses the same collision data, or a subset thereof, and shares some normalization and calibration inputs with the published analysis. However, the event-level information and full systematic covariance matrices needed to quantify these correlations are not available. We therefore use the larger of the total evaluated uncertainties from the reanalysis and the published measurement as the uncertainty scale for the comparison, rather than adding them in quadrature. This choice also accounts for bins in which the published systematic uncertainty exceeds the uncertainty evaluated in our reanalysis. The normalized deviation is defined as
\begin{equation}
\delta_i = \frac{x_i-x_i^{\mathrm{pub}}}{u_{\max,i}}
= \frac{R_i-1}{u_{\max,i}/x_i^{\mathrm{pub}}},
\qquad
R_i=\frac{x_i}{x_i^{\mathrm{pub}}},
\qquad
u_{\max,i}=\max(u_i,u_i^{\mathrm{pub}}).
\label{eq:ratioerror}
\end{equation}
where $R_i$ is the ratio of the AI-reconstructed cross section to the published CMS value. This quantity provides an approximate measure of the bin-by-bin agreement, rather than a formal statistical pull.}

In the ratio plots (Figs.~\ref{fig:cms2010ratio} and~\ref{fig:cmsHI2011promptcomparison}), the black inner error bars show the statistical uncertainty of our reanalysis, normalized to $x_i^{\mathrm{pub}}$, and the colored outer bars show its total evaluated uncertainty. The shaded gray band around $R_i=1$ represents the relative uncertainty of the published measurement. We summarize the agreement in each dataset using the mean, RMS spread, and median of $R_i$. The median provides a robust measure of an overall normalization offset, as it is less sensitive to individual outliers, particularly in sparsely populated bins near the phase-space boundaries. For the PbPb comparison, we also report the mean and RMS of $\delta_i$.

\subsection{\texorpdfstring{Inclusive $J/\psi$ Results in $pp$ Collisions at 7~TeV}{Inclusive J/psi Results in pp Collisions at 7 TeV}}
\label{sec:results_pp}

\paragraph{Mass and decay-length fits.}
Prompt and nonprompt $J/\psi$ mesons from $b$-hadron decays are separated through a simultaneous fit to the dimuon invariant mass $m_{\mu^+\mu^-}$ and pseudo-proper decay length $\ell_{J/\psi}$ (Eq.~\eqref{eq:lifetime}; Section~\ref{sec:ailifecycle}). Figure~\ref{fig:cms2010} compares the AI-reconstructed distributions with published CMS examples in the same kinematic intervals. The fitted parameters are summarized in Table~\ref{tab:cms2010fitparams}.

For the intermediate-rapidity mass-fit example ($12 < p_{\mathrm{T}} < 15~\mathrm{GeV}$, $0.9 < |y| < 1.2$; Figs.~\ref{fig:cms2010}a,b), the Crystal Ball plus Gaussian model gives an inclusive yield of $384.7\pm20.3$, a mass mean of $3.0920\pm0.0019~\mathrm{GeV}$, and a Crystal Ball core width of $34.03\pm1.56~\mathrm{MeV}$. For the midrapidity joint fit ($8 < p_{\mathrm{T}} < 9~\mathrm{GeV}$, $|y|<0.9$; Figs.~\ref{fig:cms2010}c,d), the 713-candidate sample yields $437.1\pm23.2$ prompt and $192.4\pm17.1$ nonprompt candidates. The fitted mass mean and Crystal Ball width are $3.0950\pm0.0010~\mathrm{GeV}$ and $16.34\pm0.89~\mathrm{MeV}$, respectively, while the effective nonprompt decay-length slope is $\lambda_B=0.3201\pm0.0296~\mathrm{mm}$. Both fitted mass means are consistent with the PDG value of $3.0969~\mathrm{GeV}$. The cross-section result in the joint-fit example is also compared with the published value in Table~\ref{tab:cms2010fitparams}.

\begin{table}[!htbp]
\centering\small
\caption{Fit parameters and cross-section results for the two example intervals in Fig.~\ref{fig:cms2010}, compared with published CMS results~\cite{CMS2010,CMS2010Tables}. Here $\sigma_{\mathrm{CB}}$ denotes the Crystal Ball core width and $\lambda_B$ the effective nonprompt decay-length slope in mm. Raw yields are from the $3.04~\mathrm{pb}^{-1}$ reanalysis sample; cross sections are directly comparable with the published measurement.}
\label{tab:cms2010fitparams}
\setlength{\tabcolsep}{3pt}
\begin{tabular}{@{}>{\raggedright\arraybackslash}p{0.25\linewidth}>{\raggedright\arraybackslash}p{0.23\linewidth}>{\raggedright\arraybackslash}p{0.24\linewidth}>{\raggedright\arraybackslash}p{0.24\linewidth}@{}}
\toprule
Fit parameter / observable & Mass-fit example\newline ($12 < p_{\mathrm{T}} < 15~\mathrm{GeV}$,\newline $0.9 < |y| < 1.2$) & Lifetime-fit example\newline ($8 < p_{\mathrm{T}} < 9~\mathrm{GeV}$,\newline $|y| < 0.9$) & Published CMS reference~\cite{CMS2010,CMS2010Tables} \\
\midrule
Fitted $J/\psi$ mass mean $m$ [GeV] & $3.0920 \pm 0.0019$ & $3.0950 \pm 0.0010$ & \rev{Floated per bin (no PDG mass constraint)} \\
Crystal Ball width $\sigma_{\mathrm{CB}}$ [MeV] & $34.03 \pm 1.56$ & $16.34 \pm 0.89$ & $\approx 35~\mathrm{MeV}$ ($0.9<|y|<1.2$);\newline $\approx 20~\mathrm{MeV}$ ($|y|<0.9$) \\
Nonprompt slope $\lambda_B$ [mm] & — (1D mass fit) & $0.3201 \pm 0.0296$ & Fitted per bin ($c\tau_B \approx 0.45~\mathrm{mm}$ before boost projection) \\
Fitted $J/\psi$ signal yield ($3.04~\mathrm{pb}^{-1}$) & $384.7 \pm 20.3$ (inclusive) & $437.1 \pm 23.2$ (prompt);\newline $192.4 \pm 17.1$ (nonprompt) & Full $36.7~\mathrm{pb}^{-1}$ sample ($\approx 12\times$ larger luminosity) \\
Prompt cross section $\xsec$ [nb/GeV] & \rev{— (1D mass fit; no cross section)$^*$} & $2.23 \pm 0.26~\mathrm{(stat.)} \pm 0.25~\mathrm{(syst.)}$ & \rev{—$^*$;\newline $2.54 \pm 0.07~\mathrm{(stat.)} \pm 0.12~\mathrm{(syst.)}$} \\
\bottomrule
\end{tabular}
\par\smallskip\parbox{.98\linewidth}{\footnotesize \rev{$^*$Note: The first column is a 1D inclusive mass fit and does not yield a prompt cross section; the separate 2D fit in $12 < p_{\mathrm{T}} < 15~\mathrm{GeV}$, $0.9 < |y| < 1.2$ did not pass our rate screen at $3.04~\mathrm{pb}^{-1}$ (the acceptance-corrected published CMS prompt value is $0.248 \pm 0.006 \pm 0.013 \pm 0.010~\mathrm{nb/GeV}$~\cite{CMS2010Tables} and is not used in the ratio comparison).}}
\end{table}

\paragraph{Cross-section spectrum and ratio.}
The published CMS measurement contains 44 $(p_{\mathrm{T}},|y|)$ bins~\cite{CMS2010Tables}, of which 39 are reproduced in this reanalysis. The other five bins, located near the kinematic boundaries, are excluded because of insufficient data or simulated events, or because the two-dimensional fit does not converge with a positive-definite covariance matrix. Figure~\ref{fig:cms2010rates} compares the double-differential prompt cross sections across the five rapidity intervals, while Fig.~\ref{fig:cms2010ratio} shows the corresponding ratios, $R_i=x_i/x_i^{\mathrm{pub}}$.

For the 39 reproduced bins, the ratios have a mean of $\overline{R}=1.000$, an RMS spread of $0.215$, and a median of $\mathrm{med}(R_i)=0.994$. Thirty-six bins lie within $0.68<R_i<1.31$. The other three are in the forward region ($2.1<|y|<2.4$), with ratios of $0.516\pm0.145$ ($12$--$15~\mathrm{GeV}$), $0.677\pm0.182$ ($10$--$12~\mathrm{GeV}$), and $1.678\pm0.411$ ($5.5$--$8~\mathrm{GeV}$).

In the midrapidity example bin ($8<p_{\mathrm{T}}<9~\mathrm{GeV}$, $|y|<0.9$), with an acceptance of $A_g=0.1075$, the measured prompt cross section is
$2.23\pm0.26~\mathrm{(stat.)}\pm0.05~\mathrm{(model)}\pm0.25~\mathrm{(lumi.)}~\mathrm{nb/GeV}$,
compared with the published value of
$2.54\pm0.07~\mathrm{(stat.)}\pm0.06~\mathrm{(syst.)}\pm0.10~\mathrm{(lumi.)}~\mathrm{nb/GeV}$~\cite{CMS2010Tables}.
This agreement tests the event selection, fitting, and efficiency-correction chain. The reanalysis does not, however, include $\psi(2S)$ and $\chi_c$ feed-down or pileup mixing, and lacks data-driven tag-and-probe trigger scale factors. It therefore serves as a validation of the AI-assisted workflow, rather than a replacement for the full CMS measurement and its calibration procedures.

\input{sections/compact_figures/cms2010}

\subsection{\texorpdfstring{Inclusive $J/\psi$ Results in Peripheral PbPb collisions at 2.76~TeV}{Inclusive J/psi Results in Peripheral PbPb collisions at 2.76 TeV}}
\label{sec:results_pbpb}

Table~\ref{tab:cmsHI2011corrected} and Fig.~\ref{fig:cmsHI2011promptcomparison} summarize the results in the six matched reference intervals. They present the fiducial cross sections $\sigma_{\mathrm{fid}}$, the corresponding $T_{\mathrm{AA}}$-scaled yields $\rho$ (Fig.~\ref{fig:cmsHI2011promptcomparison}a), the ratios to the published CMS values, $R_i=x_i/x_i^{\mathrm{pub}}$ (Fig.~\ref{fig:cmsHI2011promptcomparison}b), and the evaluated relative uncertainties (Fig.~\ref{fig:cmsHI2011promptcomparison}c).

The cross section and yield in the fourth interval ($1.6<|y|<2.4$, $3<p_{\mathrm{T}}<6.5~\mathrm{GeV}$) are substantially larger than those in the other intervals. This is expected from the steeply falling $J/\psi$ transverse-momentum spectrum. The relatively narrow $3<p_{\mathrm{T}}<6.5~\mathrm{GeV}$ interval covers the high-rate low-$p_{\mathrm{T}}$ region, whereas the $6.5<p_{\mathrm{T}}<30~\mathrm{GeV}$ intervals span a much wider, lower-rate region.

The ratios range from $0.78$ to $1.37$, with a mean of $1.158$, an RMS spread of $0.225$, and a median of $1.226$. The evaluated $1\sigma$ uncertainty intervals of the reanalysis overlap the published CMS uncertainty bands in all six cases. To quantify the differences, we use $\delta_i=(x_i-x_i^{\mathrm{pub}})/u_{\max,i}$, where $u_{\max,i}=\max(u_i,u_i^{\mathrm{pub}})$. Four intervals have $|\delta_i|\leq1.04$, and five have $|\delta_i|\leq1.15$. The largest deviation is $1.52$; the mean and RMS are $+0.57$ and $1.04$, respectively.

Two qualifications are important when interpreting this agreement. First, the six intervals are not statistically independent. The first three are rapidity slices of the same $50$--$100\%$ centrality sample, while the last two divide the same $|y|<2.4$, $6.5<p_{\mathrm{T}}<30~\mathrm{GeV}$ sample into two centrality ranges. They therefore represent six matched projections of overlapping data, rather than six independent datasets. Second, the heavy-ion tracking, occupancy-closure, and external tag-and-probe transfer uncertainties associated with the embedded simulation have not been independently validated. The comparison should therefore be viewed as a conditional consistency check, not an absolute validation of the heavy-ion efficiency corrections.

\input{tables/cms_hi2011_prompt_yields}
\input{tables/cms_hi2011_corrected}
\input{sections/compact_figures/cms_hi2011}

%% file: sections/compact_figures/cms2010.tex
\begin{figure}[!htbp]
\centering
\newsavebox{\CMSExampleMassPanel}
\newsavebox{\CMSExampleLifetimePanel}
\sbox{\CMSExampleMassPanel}{\includegraphics[width=.43\textwidth]{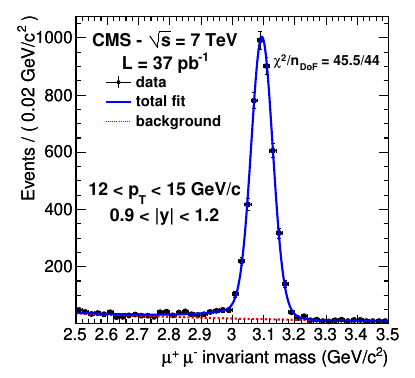}}
\sbox{\CMSExampleLifetimePanel}{\includegraphics[width=.42\textwidth]{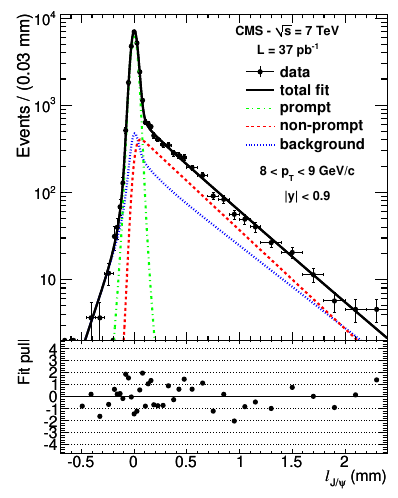}}
\begin{subfigure}[t]{.43\textwidth}\centering
\usebox{\CMSExampleMassPanel}\caption{Published mass fit ($12 < p_{\mathrm{T}} < 15~\mathrm{GeV}$, $0.9 < |y| < 1.2$).}
\end{subfigure}
\hfill
\begin{subfigure}[t]{.49\textwidth}\centering
\includegraphics[width=\linewidth,height=\ht\CMSExampleMassPanel,keepaspectratio]{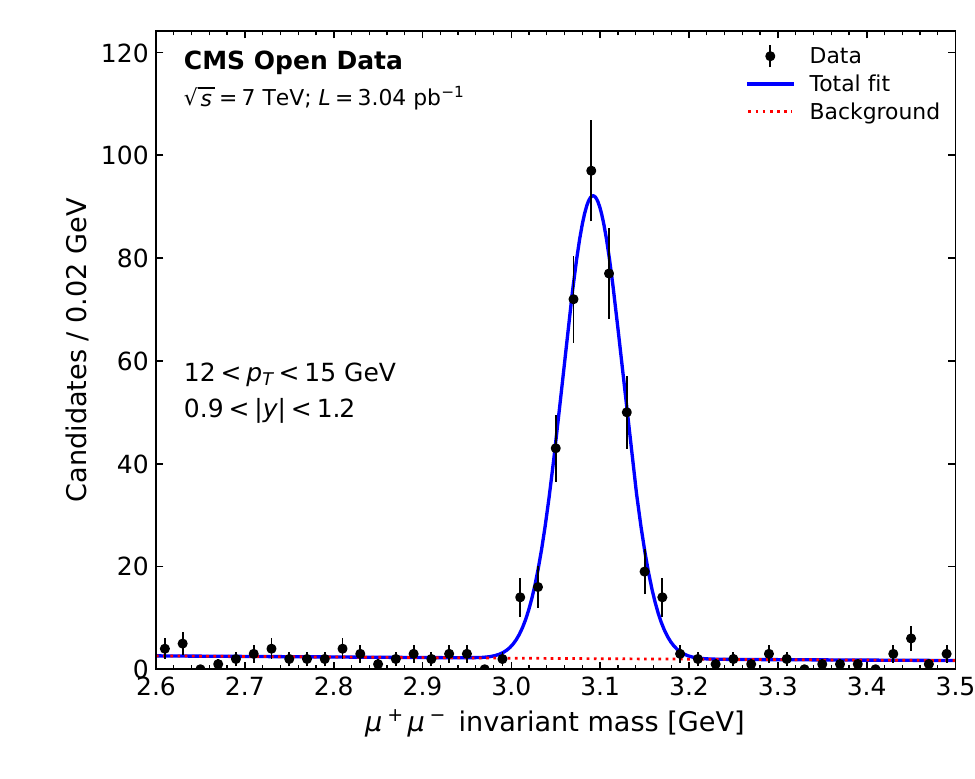}\caption{AI reanalysis 1D mass fit ($12 < p_{\mathrm{T}} < 15~\mathrm{GeV}$, $0.9 < |y| < 1.2$).}
\end{subfigure}\\[1.5mm]
\begin{subfigure}[t]{.42\textwidth}\centering
\usebox{\CMSExampleLifetimePanel}\caption{Published lifetime fit ($8 < p_{\mathrm{T}} < 9~\mathrm{GeV}$, $|y| < 0.9$).}
\end{subfigure}
\hfill
\begin{subfigure}[t]{.49\textwidth}\centering
\includegraphics[width=\linewidth,height=\ht\CMSExampleLifetimePanel,keepaspectratio]{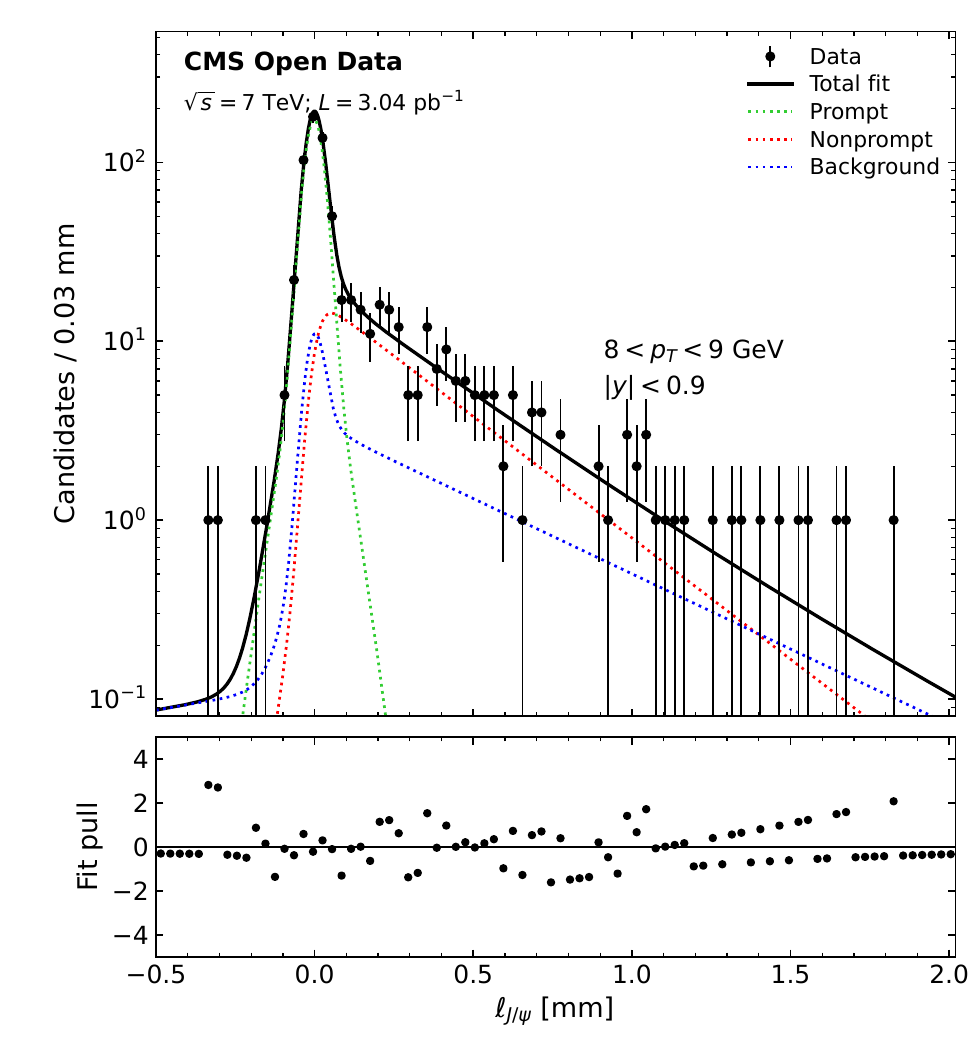}\caption{AI reanalysis lifetime fit ($8 < p_{\mathrm{T}} < 9~\mathrm{GeV}$, $|y| < 0.9$).}
\end{subfigure}
\caption{CMS 2010 dimuon invariant-mass (top row, $12 < p_{\mathrm{T}} < 15~\mathrm{GeV}$, $0.9 < |y| < 1.2$) and pseudo-proper decay-length $\ell_{J/\psi}$ (bottom row, $8 < p_{\mathrm{T}} < 9~\mathrm{GeV}$, $|y| < 0.9$) fit projections comparing the published full-sample panels ($36.7~\mathrm{pb}^{-1}$, left)~\cite{CMS2010} with the AI-assisted reanalysis of the unprescaled Tight double-muon trigger open-data subset ($3.04~\mathrm{pb}^{-1}$, right). Panel~(b) is a one-dimensional inclusive mass fit shown for shape illustration only: the two-dimensional fit in this bin did not pass the rate screen, so the bin is excluded from the reanalysis cross-section and ratio comparisons. Quantitative fit parameters and uncertainties are listed in Table~\ref{tab:cms2010fitparams}.}\label{fig:cms2010}
\end{figure}
\begin{figure}[!htbp]
\centering
\begin{subfigure}[t]{.43\textwidth}\centering\includegraphics[width=\linewidth]{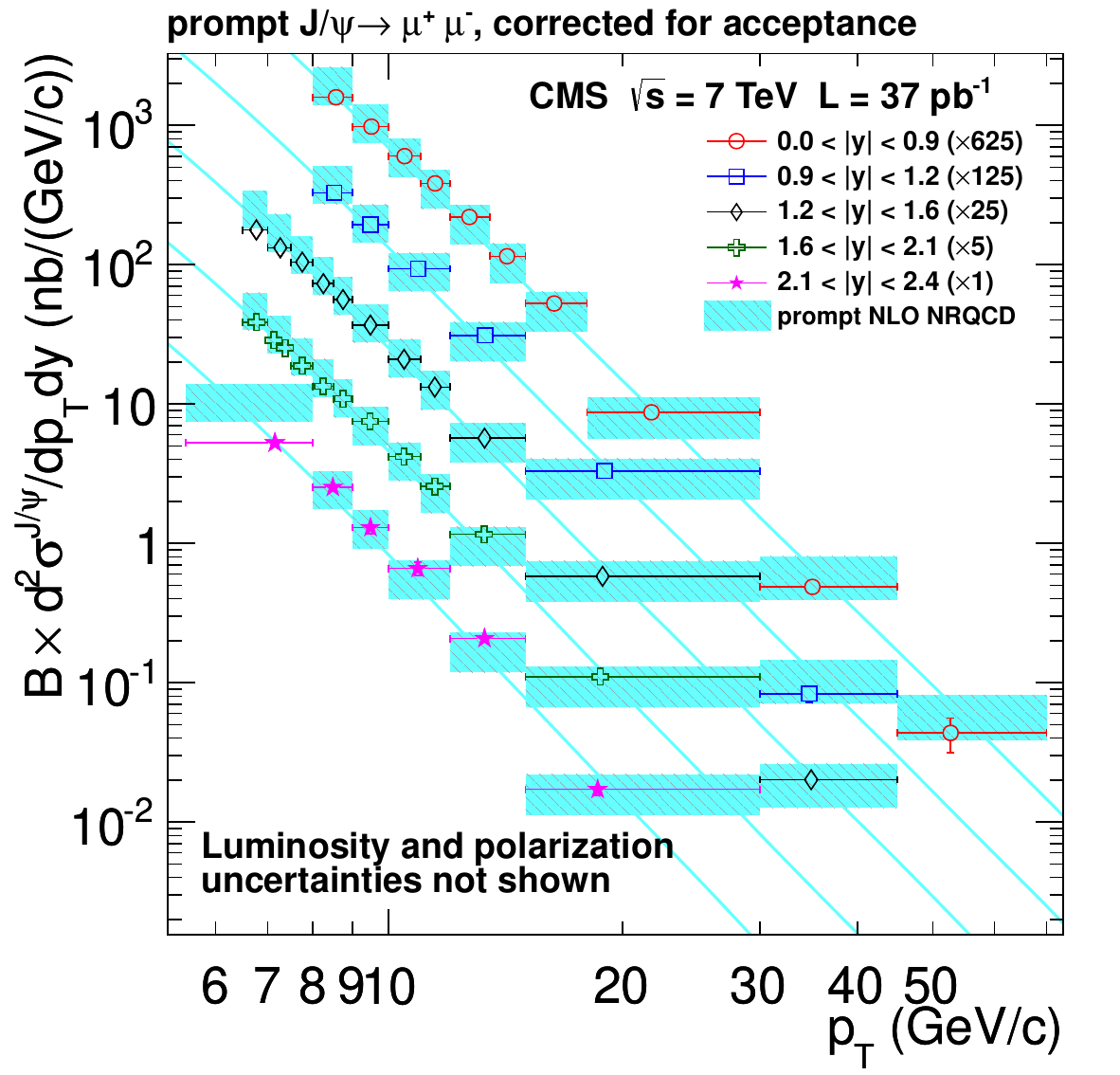}\caption{Published CMS prompt $J/\psi$ cross sections ($36.7~\mathrm{pb}^{-1}$).}\end{subfigure}
\hfill
\begin{subfigure}[t]{.49\textwidth}\centering\includegraphics[width=\linewidth]{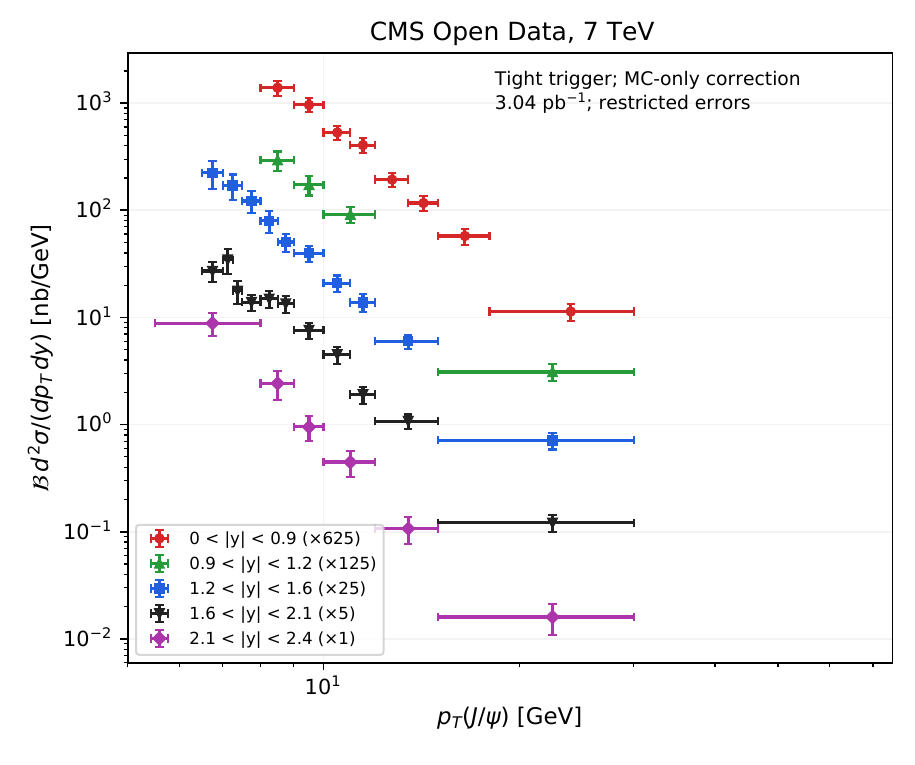}\caption{AI-assisted reanalysis ($3.04~\mathrm{pb}^{-1}$ Tight trigger).}\end{subfigure}
\caption{CMS 2010 published (left)~\cite{CMS2010Tables} and AI-reconstructed (right) prompt $J/\psi$ double-differential cross sections \rev{times dimuon branching fraction, $\mathcal{B}(J/\psi\to\mu^+\mu^-)\,d^2\sigma/(dp_{\mathrm{T}}\,dy)$} (in nb/GeV), across five $|y|$ slices (scaled by factors of 625, 125, 25, 5, and 1 for visibility) using identical $(p_{\mathrm{T}}, |y|)$ bin boundaries.}\label{fig:cms2010rates}
\end{figure}
\begin{figure}[!htbp]\centering
\includegraphics[width=.84\linewidth]{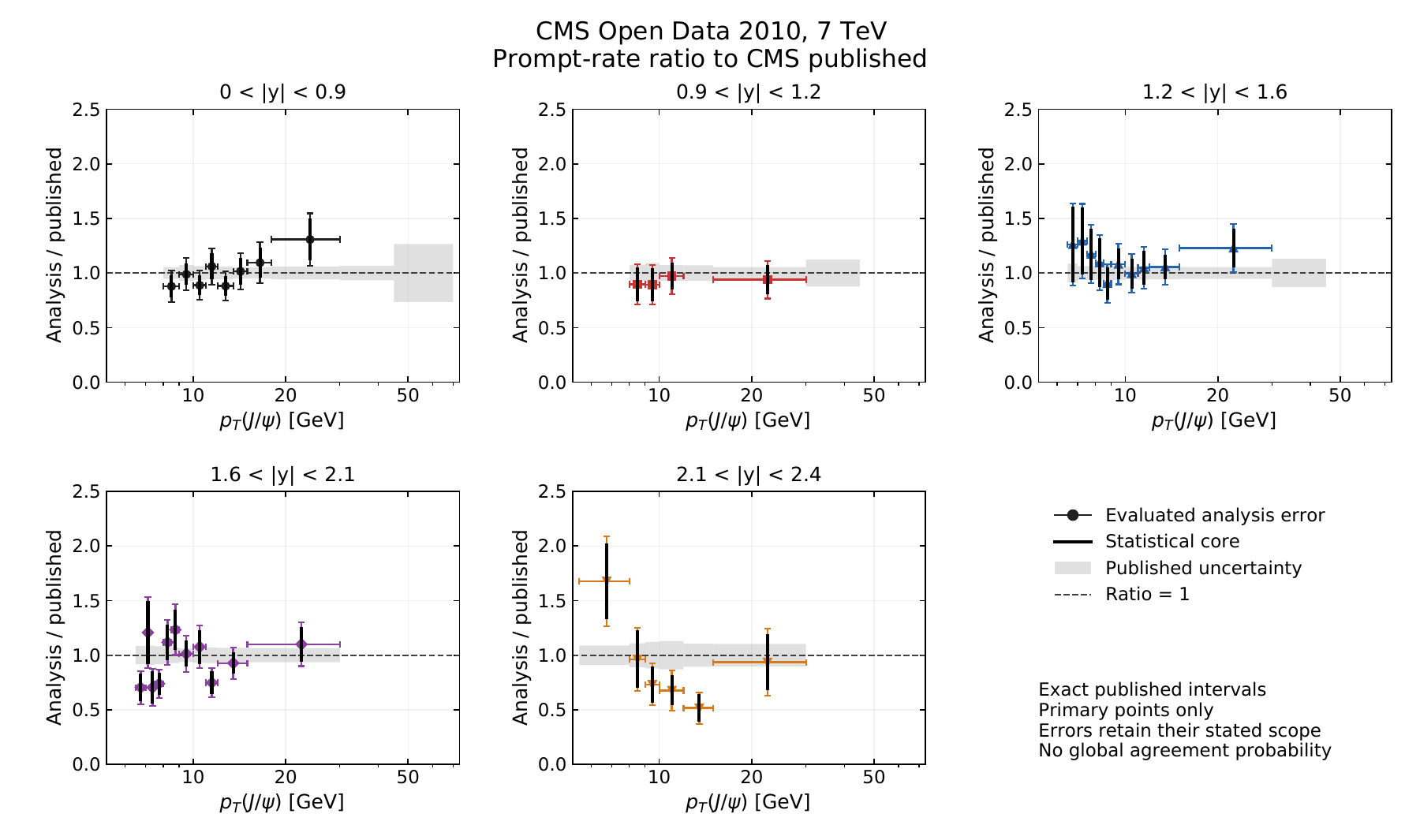}
\caption{CMS 2010 bin-by-bin ratio \rev{$R_i = x_i / x_i^{\mathrm{pub}}$} of the AI-assisted reanalysis to the published CMS prompt $J/\psi$ cross section across the 39 populated $(p_{\mathrm{T}}, |y|)$ bins~\cite{CMS2010Tables}, showing a mean ratio of $1.000$ with an RMS spread of $0.215$ and a median ratio of $0.994$. Error bars combine the evaluated statistical, simulation, fit-model, and luminosity uncertainties.}\label{fig:cms2010ratio}
\end{figure}

%% file: tables/cms_hi2011_prompt_yields.tex
\begin{table}[!htbp]\centering\small
\caption{Fitted prompt $J/\psi$ signal yields in the six peripheral ($50$--$100\%$ centrality) PbPb kinematic and centrality intervals of Ref.~\cite{CMSHI2011Jpsi}, showing the total number of selected dimuon pairs, the fitted prompt yield with its statistical uncertainty, the fit-model variation envelope, and the published CMS \rev{$T_{\rm AA}$-scaled differential prompt $J/\psi$ yield per unit transverse momentum, $\rho^{\mathrm{pub}} \equiv \mathcal{B}(J/\psi\to\mu^+\mu^-)\,(dN_{\mathrm{PbPb}}/dp_{\mathrm{T}})/\langle T_{\mathrm{AA}}\rangle$} (in pb/GeV, integrated over the stated $|y|$ interval, with statistical and systematic uncertainties)~\cite{CMSHI2011Jpsi}. \rev{In the lowest-multiplicity $60$--$100\%$ bin (marked by $\dagger$), the unphysical negative-decay-length background component vanished in the fit and was fixed to zero so that the covariance matrix remained well-defined, yielding asymmetric profile-likelihood statistical uncertainties ($+18.2 / -17.6$ events).}}\label{tab:cmsHI2011prompt}
\begin{tabular}{ccccrrl}\toprule Centrality [\%] & $|y|$ & $p_{\mathrm{T}}$ [GeV] & Selected pairs & Prompt $J/\psi$ yield & Model envelope & Published $\rho^{\mathrm{pub}}$ [pb/GeV]\\\midrule
50--100 & 0--1.2 & 6.5--30 & 469 & $283.7\pm18.5$ & $\pm9.5$ & $47\pm3\pm8$\\
50--100 & 1.2--1.6 & 6.5--30 & 226 & $139.4\pm13.8$ & $\pm7.0$ & $71\pm6\pm12$\\
50--100 & 1.6--2.4 & 6.5--30 & 293 & $169.3\pm16.1$ & $\pm4.3$ & $46\pm4\pm8$\\
50--100 & 1.6--2.4 & 3--6.5 & 637 & $353.3\pm24.5$ & $\pm13.3$ & $815\pm53\pm158$\\
50--60 & 0--2.4 & 6.5--30 & 556 & $334.2\pm21.7$ & $\pm14.8$ & $50\pm3\pm8$\\
60--100 & 0--2.4 & 6.5--30 & 432 & $256.0^{+18.2}_{-17.6}\,\dagger$ & $\pm7.2$ & $50\pm3\pm9$\\
\bottomrule\end{tabular}
\par\vspace{0.5mm}\parbox{.98\linewidth}{\footnotesize \rev{Note: The first three rows subdivide the $50$--$100\%$ peripheral sample into three rapidity slices at $6.5<p_{\mathrm{T}}<30~\mathrm{GeV}$ ($469+226+293=988$ pairs), the last two rows subdivide the same $|y|<2.4$, $6.5<p_{\mathrm{T}}<30~\mathrm{GeV}$ sample into $50$--$60\%$ and $60$--$100\%$ centrality ($556+432=988$ pairs), and the fourth row contains $637$ additional candidate pairs in the disjoint low-$p_{\mathrm{T}}$ interval ($3<p_{\mathrm{T}}<6.5~\mathrm{GeV}$). The first three and last two rows are two overlapping projections of the same 988 high-$p_{\mathrm{T}}$ pairs; the fourth row adds 637 disjoint low-$p_{\mathrm{T}}$ pairs. The six rows therefore contain 1,625 unique pairs, and their sum (2,613) is not a unique candidate count.}}\vspace{-1mm}
\end{table}

%% file: tables/cms_hi2011_corrected.tex
\begin{table}[!htbp]\centering\small
\caption{Efficiency-corrected peripheral prompt $J/\psi$ results across the six published CMS reference intervals~\cite{CMSHI2011Jpsi}: the inverse acceptance-times-efficiency correction factor $C$ (with shared-background Monte Carlo statistical uncertainty), \rev{the differential fiducial cross section per unit transverse momentum $\sigma_{\rm fid} \equiv \mathcal{B}(J/\psi\to\mu^+\mu^-)\,d\sigma_{\rm fid}/dp_{\mathrm{T}}$ (in nb/GeV, integrated over the stated $|y|$ interval), the $T_{\rm AA}$-scaled differential fiducial yield per unit transverse momentum $\rho \equiv \mathcal{B}(J/\psi\to\mu^+\mu^-)\,(dN_{\mathrm{PbPb}}/dp_{\mathrm{T}})/\langle T_{\mathrm{AA}}\rangle$ (in pb/GeV), and the ratio $R_i = \rho_i / \rho_i^{\mathrm{pub}}$ of our reanalysis to the published CMS value}. For $\sigma_{\rm fid}$, $\rho$, and $R_i$, \rev{the quoted uncertainties are absolute uncertainties}: the first combines data-fit, Monte Carlo, and single-muon calibration statistical uncertainties, and the second is the fit-model variation envelope.}\label{tab:cmsHI2011corrected}
\begin{tabular}{cccrrrr}\toprule Centrality [\%] & $|y|$ & $p_{\mathrm{T}}$ [GeV] & Correction $C$ & $\sigma_{\rm fid}$ [nb/GeV] & $\rho$ [pb/GeV] & \rev{Reanalysis / published $R_i$}\\\midrule
50--100 & 0--1.2 & 6.5--30 & $2.49\pm0.35$ & $111.5\pm17.7\pm3.7$ & $62.3\pm9.9\pm2.1$ & $1.33\pm0.21\pm0.04$\\
50--100 & 1.2--1.6 & 6.5--30 & $2.64\pm0.57$ & $174.0\pm44.3\pm8.7$ & $97.2\pm24.7\pm4.9$ & $1.37\pm0.35\pm0.07$\\
50--100 & 1.6--2.4 & 6.5--30 & $1.59\pm0.23$ & $63.9\pm12.7\pm1.6$ & $35.7\pm7.1\pm0.9$ & $0.78\pm0.15\pm0.02$\\
50--100 & 1.6--2.4 & 3--6.5 & $3.34\pm0.32$ & $1875.5\pm405.4\pm70.6$ & $1047.7\pm226.5\pm39.4$ & $1.29\pm0.28\pm0.05$\\
50--60 & 0--2.4 & 6.5--30 & $2.02\pm0.26$ & $53.3\pm8.6\pm2.4$ & $51.2\pm8.3\pm2.3$ & $1.02\pm0.17\pm0.05$\\
60--100 & 0--2.4 & 6.5--30 & $2.17\pm0.22$ & $43.9\pm6.3\pm1.2$ & $58.3\pm8.3\pm1.6$ & $1.17\pm0.17\pm0.03$\\
\bottomrule\end{tabular}
\par\vspace{0.5mm}\parbox{.98\linewidth}{\footnotesize Note: Published single-muon tag-and-probe efficiency parameterizations~\cite{CMSHI2011Calibration} are applied to the embedded simulation; additional heavy-ion tracking, occupancy-closure, and Glauber-normalization uncertainties are not included in the evaluated error bars. \rev{In the fourth row ($3<p_{\mathrm{T}}<6.5~\mathrm{GeV}$), $\sigma_{\rm fid}$ and $\rho$ are an order of magnitude larger than in the $6.5<p_{\mathrm{T}}<30~\mathrm{GeV}$ rows because the $d\sigma/dp_{\mathrm{T}}$ spectrum peaks at low $p_{\mathrm{T}}$ (matching $\rho^{\mathrm{pub}} = 815\pm53\pm158~\mathrm{pb/GeV}$ in Table~\ref{tab:cmsHI2011prompt}).}}\vspace{-1mm}
\end{table}

%% file: sections/compact_figures/cms_hi2011.tex
\begin{figure}[H]\centering
\includegraphics[width=.49\linewidth,height=52mm,keepaspectratio]{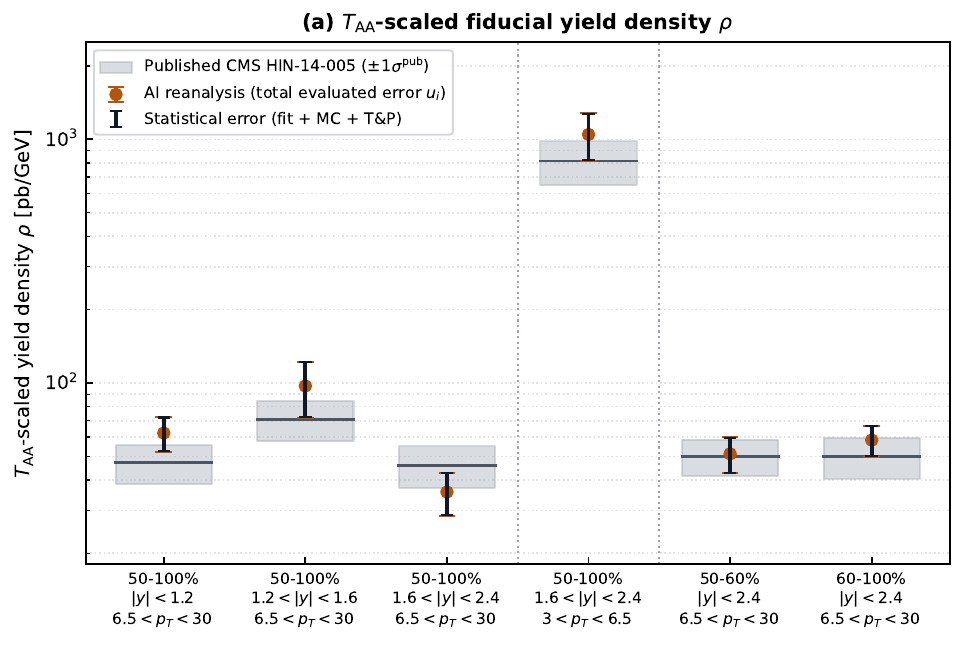}\hfill
\includegraphics[width=.49\linewidth,height=52mm,keepaspectratio]{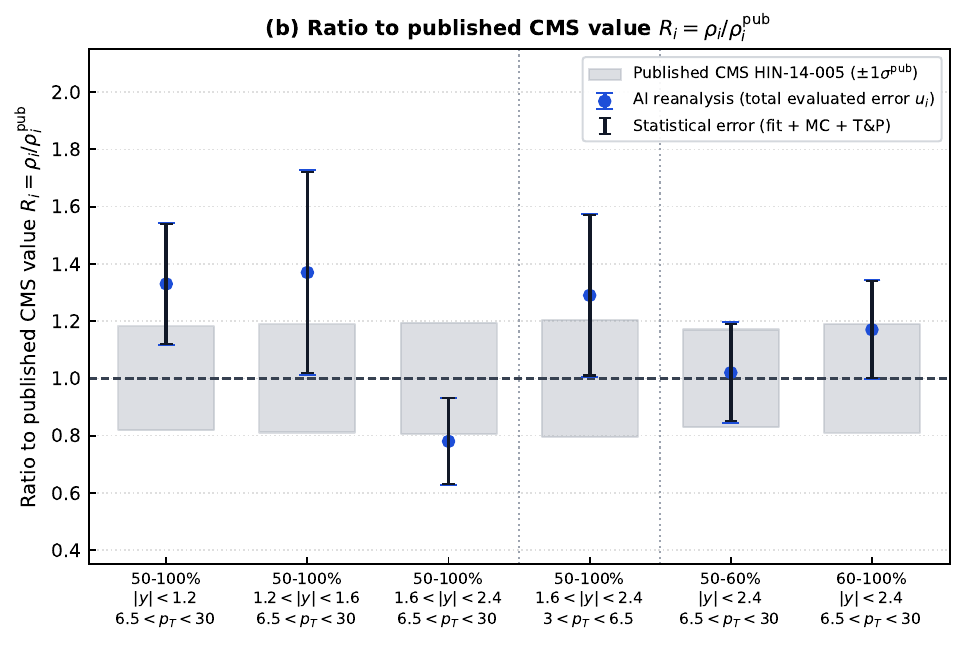}\\[1.5mm]
\includegraphics[width=.54\linewidth,height=50mm,keepaspectratio]{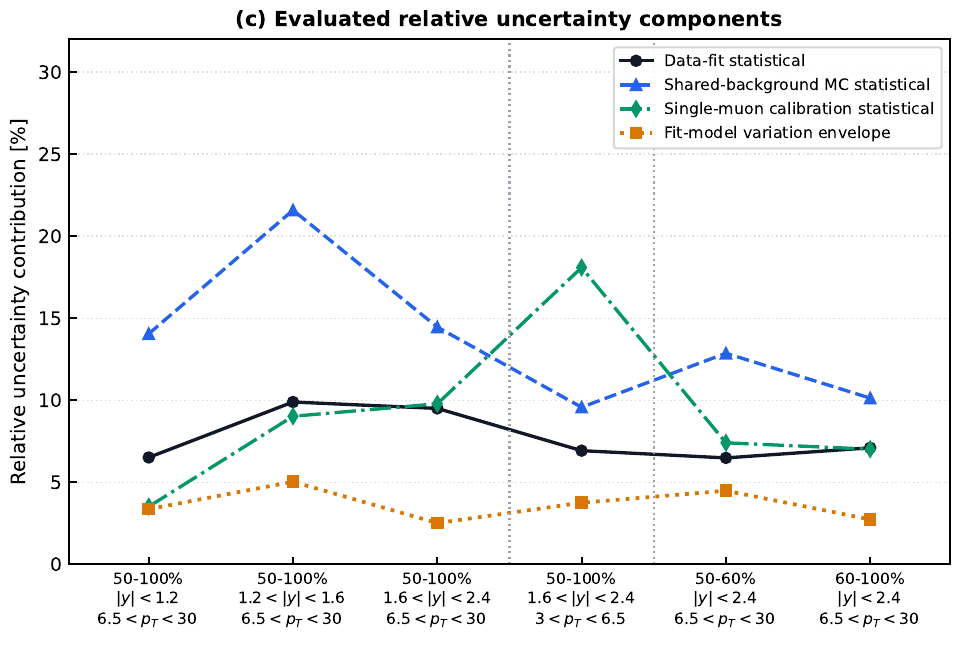}
\caption{CMS PbPb 2011 peripheral prompt $J/\psi$ results \rev{displayed side by side across all six matched reference intervals of Ref.~\cite{CMSHI2011Jpsi} (vertical dotted lines separate the three $50$--$100\%$ high-$p_{\mathrm{T}}$ rapidity slices, the forward low-$p_{\mathrm{T}}$ interval, and the two $|y|<2.4$ centrality sub-intervals $50$--$60\%$ and $60$--$100\%$, which analyze the same collision events as the first three high-$p_{\mathrm{T}}$ slices)}: \revb{(a)~$T_{\rm AA}$-scaled differential fiducial yield $\rho$ (upper left; the proportional unscaled fiducial cross section $\sigma_{\mathrm{fid}}$ is listed in Table~\ref{tab:cmsHI2011corrected}), (b)}~\rev{ratio $R_i = \rho_i / \rho_i^{\mathrm{pub}}$ of our reanalysis to the published CMS value} (\revb{upper right}, mean $1.158$, RMS $0.225$, median $1.226$; \rev{the evaluated $1\sigma$ reanalysis error bars overlap the published $1\sigma$ bands in all six intervals}), and \revb{(c)~evaluated relative uncertainty components (bottom)}. Black inner bars combine data-fit, shared-background MC, and calibration statistical errors; colored outer bars add the fit-model envelope, and gray bands denote the published uncertainties.}\label{fig:cmsHI2011promptcomparison}
\end{figure}

%% file: sections/limitations.tex
\section{Pitfalls and challenges: why human oversight is indispensable}
\label{sec:discussion}

\begin{figure}[H]
\centering
\includegraphics[width=0.96\linewidth]{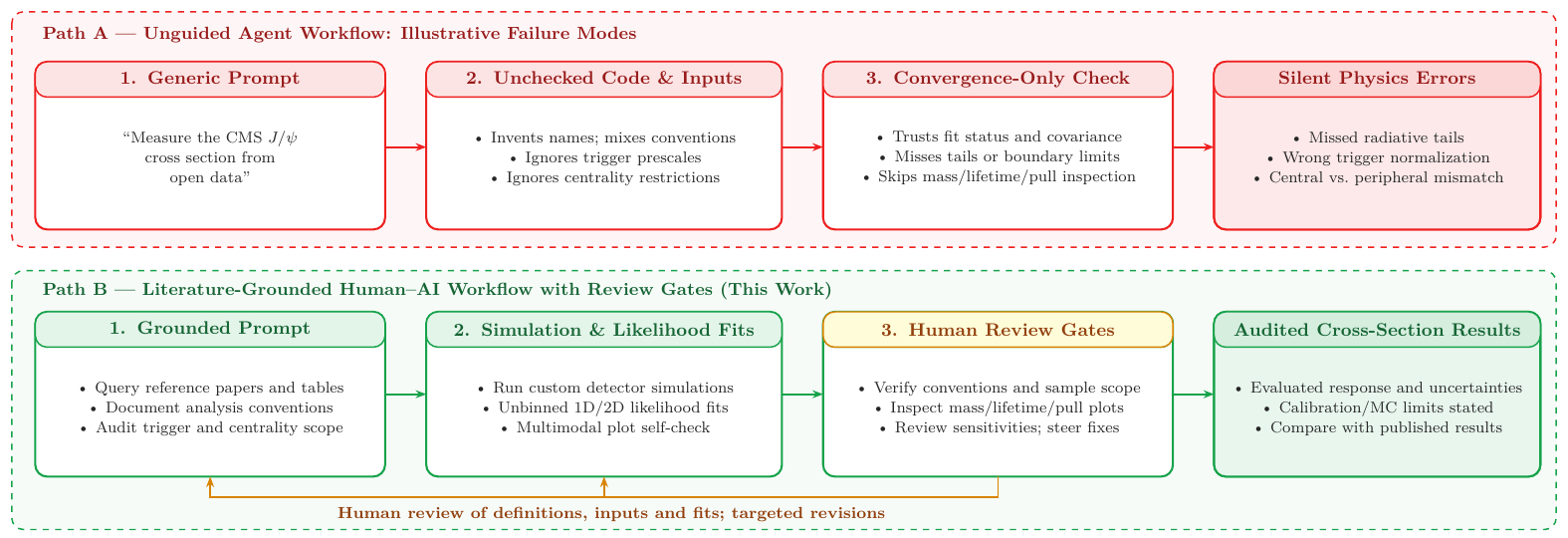}
\caption{Schematic illustration of agent failure modes and the safeguards used in this study. Path~A illustrates unchecked analysis conventions and input scope, and over-reliance on numerical convergence. Path~B combines literature-based instructions with human review of definitions, inputs, fit projections, and model sensitivity. Its reported response and uncertainty budget cover the evaluated contributions; remaining calibration and simulation limitations are stated explicitly.}
\label{fig:aigate}
\end{figure}

\paragraph{Failure modes of the agent.}
The two analyses revealed three recurring failure modes that distinguish LLM-based coding agents from conventional programming tools (Fig.~\ref{fig:aigate}):

\begin{itemize}
\setlength{\itemsep}{0.5pt}\setlength{\parskip}{0pt}
\item \textbf{Hallucinations and silent convention drift:}
Even with access to a local reference archive, an LLM agent can invent variable names or silently apply conventions from one analysis to another. One example is using the wrong decay-length reference: the 2010 $pp$ analysis uses the transverse beam-spot position, while the 2011 PbPb analysis uses the event-by-event primary vertex (Section~\ref{sec:observables}). Other examples include changing rapidity or centrality definitions or the convention for including the dimuon branching fraction. Requiring the agent to document these conventions from the reference papers before coding, followed by human verification, helps prevent such errors.

\item \textbf{Over-reliance on numerical convergence:}
In batch fitting, the agent tends to judge fit quality using numerical indicators, such as the Minuit status and covariance-matrix quality, and may declare a fit successful without inspecting the distributions. Even with multimodal vision, it can miss physical defects that are apparent to an experienced physicist, including an inadequately modeled radiative mass tail, parameters pinned at boundary limits, or a mis-scaled background component on a logarithmic lifetime axis. Automated convergence checks must therefore be complemented by human inspection of mass, lifetime, and pull distributions at each review checkpoint.

\item \textbf{Uncritical acceptance of archive metadata and trigger configurations:}
An agent may process any input file that opens successfully without checking whether prescaled triggers bias the effective luminosity or whether reprocessing filters restrict the available phase space. For example, the $50$--$100\%$ peripheral PbPb sample cannot be directly compared with a published result for $0$--$20\%$ central collisions. As discussed in Section~\ref{sec:2012case}, human-guided audits and dedicated simulation were needed to identify and address these constraints.
\end{itemize}

\begingroup
\paragraph{Typical errors identified in the agent sessions.}
After completing the analyses, we reviewed the retained Codex session transcripts and code-execution logs to assemble a documented, non-exhaustive sample of 64 AI errors. Each event was assigned to one primary category; repeated discussions of the same error were merged, and routine debugging, informative failed tests and external outages were excluded. Table~\ref{tab:aierrors} reports the counts and percentages within this sample.
\endgroup

\input{tables/ai_error_examples}

These audits had direct consequences for the analyses. The 2011 PbPb audit identified the $50$--$100\%$ peripheral centrality restriction before an invalid comparison with central-collision results could be made (Section~\ref{sec:cmsHI2011}). For the 2010 $pp$ analysis, it identified an unprescaled trigger path that could be consistently emulated in the detector simulation (Section~\ref{sec:cms2010results}).

\paragraph{Areas for further improvement.}
The AI-assisted workflow can be further improved through more structured task instructions, systematic reference-based validation, and more efficient human oversight. Future studies could also benchmark the workflow against conventional manual analyses to quantitatively evaluate potential gains in efficiency and accuracy.

Further improvements in physics precision could be achieved with more complete calibration and simulation inputs. For the 2010 $pp$ analysis, these include data-to-simulation trigger scale factors, pileup mixing, and $\psi(2S)$ and $\chi_c$ feed-down corrections. For the 2011 PbPb analysis, additional validation of the external single-muon efficiency calibrations applied to the embedded simulation would strengthen the results. Such improvements would benefit from access to dedicated single-muon control-trigger streams, minimum-bias pileup libraries, and detailed detector-alignment systematic information, which are not included in the selected dimuon open-data releases.

\FloatBarrier

%% file: tables/ai_error_examples.tex
\begin{table}[H]
\centering
\begingroup
\captionsetup{font={footnotesize,color=black},labelfont={bf,color=black}}
\caption{Retrospective error sample from the broader workflow development ($N=64$). Colors and letters identify the same categories in the table and chart.}
\label{tab:aierrors}
\footnotesize
\begin{minipage}[c]{0.57\linewidth}
\begin{tabularx}{\linewidth}{@{}lXrr@{}}
\toprule
& Error type & Count & Share (\%) \\
\midrule
\textcolor[HTML]{8BB9D7}{\rule{5pt}{5pt}}\,A & Definition, input or reference mismatch & 21 & 32.8 \\
\textcolor[HTML]{D0E3EF}{\rule{5pt}{5pt}}\,B & Figure and document errors & 17 & 26.6 \\
\textcolor[HTML]{285A8E}{\rule{5pt}{5pt}}\,C & Task drift & 5 & 7.8 \\
\textcolor[HTML]{3D78A8}{\rule{5pt}{5pt}}\,D & Premature stopping & 5 & 7.8 \\
\textcolor[HTML]{619BC2}{\rule{5pt}{5pt}}\,E & Insufficient scientific validation & 5 & 7.8 \\
\textcolor[HTML]{A9CDE1}{\rule{5pt}{5pt}}\,F & Lost corrections or requirements & 5 & 7.8 \\
\textcolor[HTML]{183E6C}{\rule{5pt}{5pt}}\,G & Unsupported assertions & 4 & 6.3 \\
\textcolor[HTML]{EBF3F8}{\rule{5pt}{5pt}}\,H & Other & 2 & 3.1 \\
\midrule
& Total & 64 & 100.0 \\
\bottomrule
\end{tabularx}
\end{minipage}\hfill
\begin{minipage}[c]{0.40\linewidth}
\centering
\includegraphics[width=\linewidth]{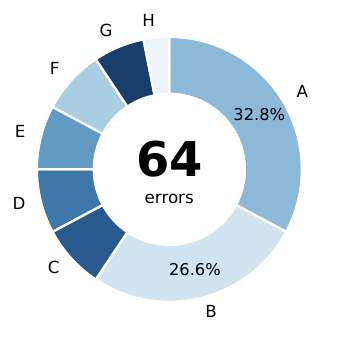}
\end{minipage}
\endgroup
\end{table}

%% file: sections/conclusions.tex
\section{Summary}
\label{sec:conclusions}

\rev{We investigated the use of a large-language-model coding agent (OpenAI's Codex, powered by the GPT-6.1 Sol and GPT-6 Astra models) in collider physics analyses under human supervision. Using charmonium production in CMS Open Data as end-to-end test cases, the workflow reproduced the published prompt $J/\psi$ differential cross sections in 39 kinematic bins of the 2010 $\sqrt{s}=7~\mathrm{TeV}$ $pp$ dataset, with a mean ratio of $1.000$, an RMS of $0.215$, and a median of $0.994$. For the 2011 $\sqrt{s_{\mathrm{NN}}}=2.76~\mathrm{TeV}$ peripheral ($50$--$100\%$ centrality) PbPb dataset, it achieved conditional agreement across six matched reference intervals, with a mean ratio of $1.158$, an RMS of $0.225$, and a median of $1.226$. The reanalysis uncertainty intervals overlap the published CMS uncertainties in all six intervals.}

\rev{The agent handled much of the software development and execution, including event-tree inspection, C++/ROOT event selection, likelihood fitting, systematic model variations, and custom detector simulations using Pythia6, HYDJET, and Geant4. Reliable results nevertheless depended on three safeguards: a local archive of reference papers and experimental records; a staged workflow with mandatory human review of numerical and visual fit diagnostics; and explicit accounting of systematic uncertainties, with unvalidated effects documented rather than assumed to be negligible. These safeguards address the failure modes identified in Section~\ref{sec:discussion} and demonstrate that human oversight remains essential for physics validation. Despite these challenges, the AI agent substantially accelerated the analysis workflow and reduced manual effort.}

Building on this workflow, we are extending the approach to additional public collider datasets. The goal is to move beyond reproducing established measurements and investigate kinematic regions and observables not covered by the original publications.

%% file: sections/acknowledgements.tex
\section*{Acknowledgements}
We thank the CMS Collaboration and the CERN Open Data team for providing the public data, open-source software, simulation and documentation used in this study. We also thank CERN for providing the lxplus computing infrastructure on which all physics analyses were performed.

AI agents using OpenAI's GPT-6 Astra and GPT-6.1 Sol assisted with analysis software development, diagnostic checks, figure preparation and manuscript editing. Responsibility for the analysis, interpretation and conclusions remains with the human researchers.

This work is partially supported by the Natural Science Foundation of China under Grants No.~12535004 and No.~W2443001, as well as the Ministry of Science and Technology of China under Grants No.~2023YFA1605804. \\ \\